\documentclass[aps,pra,reprint]{revtex4-1}
\usepackage{blindtext}
\usepackage{amssymb}
\usepackage{amsmath}
\usepackage{tikz}
\usepackage{graphicx}
\usepackage{mathrsfs}
\graphicspath{{figures/}}
\usepackage{siunitx}

\newtheorem{prop}{Proposition}\def\PRO{\begin{prop}}\def\ORP{\end{prop}}
\newtheorem{coro}{Corollary}\def\COR{\begin{coro}}\def\ROC{\end{coro}}
\newtheorem{theo}{Theorem}\def\TH{\begin{theo}}\def\HT{\end{theo}}
\def\TH{\begin{theo}}\def\HT{\end{theo}}
\newtheorem{defi}[prop]{Definition}\def\DE{\begin{defi}}\def\ED{\end{defi}}

\newtheorem{lemme}[prop]{Lemma}\def\LE{\begin{lemme}}\def\EL{\end{lemme}}
\usepackage{color}
\usepackage{graphicx,float}
\usepackage{epsfig}
\usepackage{amsmath,amssymb}
\usepackage{bm}
\usepackage[latin1]{inputenc}
\usepackage{hyphenat}
\usepackage[bookmarks=false]{hyperref}
\usepackage{color}
\usepackage[capitalise]{cleveref}
\usepackage{epstopdf}
\usepackage{braket}
\usepackage{verbatim}
\usepackage{comment}
\usepackage{multirow}
\usepackage{float}

\newcommand{\beq}{\begin{equation}}
\newcommand{\eeq}{\end{equation}}

\definecolor{pink}{RGB}{255,0,255}

\definecolor{ss_color}{rgb}{0,0,1}

\definecolor{darkorange}{RGB}{255,120,0}

\definecolor{red}{rgb}{1,0,0}

\begin{document}

\title{Foiling covert channels and malicious classical post-processing units in quantum key distribution}
\author{Marcos Curty$^1$}
\email{mcurty@com.uvigo.es}
\author{Hoi-Kwong Lo$^{2}$}
\affiliation{$^1$Escuela de Ingenier\'ia de Telecomunicaci\'on, Department of Signal Theory and Communications, University of Vigo, Vigo E-36310, Spain\\
$^2$Center for Quantum Information and Quantum Control, Department of Physics and Department of Electrical \& Computer Engineering, University of Toronto, M5S 3G4 Toronto, Canada}

\date{\today}

\begin{abstract}
Existing security proofs of quantum key distribution (QKD) suffer from two fundamental weaknesses. First, memory attacks have emerged as an important threat to the security of even device-independent quantum key distribution (DI-QKD), whenever QKD devices are re-used. This type of attacks constitutes an example of covert channels, which have attracted a lot of attention in security research in conventional cryptographic and communication systems. Second, it is often implicitly assumed that the classical post-processing units of a QKD system are trusted. This is a rather strong assumption and is very hard to justify in practice. Here, we propose a simple solution to these two fundamental problems. Specifically, we show that by using verifiable secret sharing and multiple optical devices and classical post-processing units, one could re-establish the security of QKD. Our techniques are rather general and they apply to both DI-QKD and non-DI-QKD.
\end{abstract}

\maketitle

\section{Introduction}

There has been much interest in the subject of quantum key distribution (QKD) in recent years because it holds the promise of providing information-theoretically secure communications based on the laws of quantum physics~\cite{qkd1,qkd2,qkd3}. There is, however, a big gap between the theory~\cite{qkd_t1,qkd_t2} and the practice~\cite{exp1,exp2,exp3,exp4,exp5} of QKD, and the security of QKD implementations is seriously threatened by quantum hacking~\cite{hack4,hack5,hack6,hack7,hack9}. To solve this problem, the ultimate solution is device-independent (DI)-QKD~\cite{diQKD1,diQKD2,diQKD5,diQKD6}, whose security is essentially based on a loophole-free Bell test~\cite{bell1,bell2}. Although no experimental implementation of DI-QKD has been realised yet, the recent demonstrations of loophole-free Bell tests~\cite{belltest1,belltest2,belltest3,belltest4,belltest5} might bring DI-QKD closer to experimental realisation.

\begin{figure}
  \includegraphics[width=0.95\columnwidth]{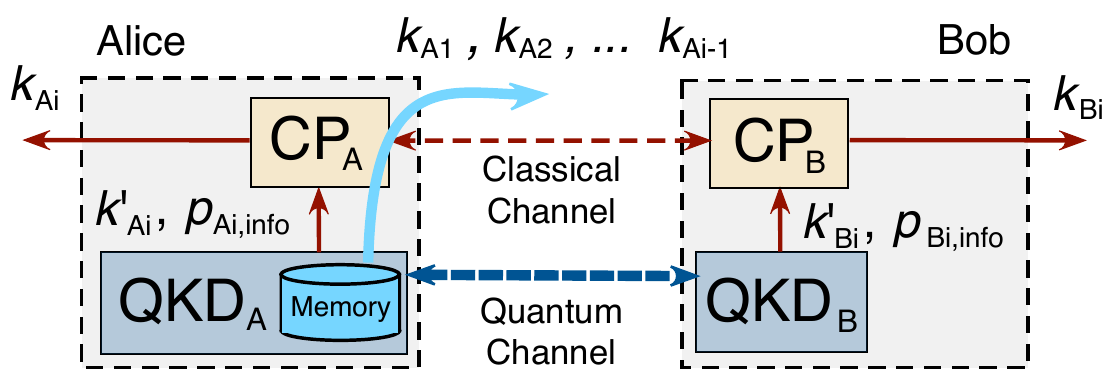}
\caption{Schematic representation of a memory attack against a DI-QKD system~\cite{mem_attack}.~In the $i$-th QKD run, Alice and Bob's QKD modules, QKD$_{\rm A}$ and QKD$_{\rm B}$, use a quantum channel to produce a raw key, $k'_{{\rm A}i}$ and $k'_{{\rm B}i}$, as well as certain protocol information, $p_{\rm A{\it i}, info}$ and $p_{\rm B{\it i}, info}$, respectively. The content of $p_{\rm A{\it i},info}$ and $p_{\rm B{\it i},info}$ depends on the particular QKD protocol implemented. For instance, in the standard decoy-state BB84 scheme~\cite{decoy1,decoy2,decoy3}, $p_{\rm A,info}$ contains the basis and the decoy setting information per emitted signal, while $p_{\rm B,info}$ contains Bob's measurement basis and it also indicates which signals produced a click in his measurement apparatus. The raw key and the protocol information is sent to Alice and Bob's classical post-processing units, CP$_{\rm A}$ and CP$_{\rm B}$, which are connected via an authenticated classical channel. These units generate a secret key, $k_{{\rm A}i}$ and $k_{{\rm B}i}$, by applying various post-processing techniques. In a memory attack, Eve hides a memory in say Alice's module QKD$_{\rm A}$ to first store up the key material generated in each QKD run and then leak this information to her by hiding it in say the decision of abort or not abort of a subsequent QKD session. For example, if a particular bit value, say the $j$-th bit, of the key generated in the first QKD run is $0$, then this memory makes that a permuted $\sigma(j)$-th QKD run aborts. (Here, $\sigma$ is a permutation and $\sigma(j)$ is the permuted value of $j$.) This could be achieved, for instance, by outputting a raw key with a high quantum bit error rate (QBER). And, if the $j$-th bit value of the key is $1$, then the $\sigma(j) $-th QKD run does not abort. That is, by simply learning whether or not the $\sigma(j)$-th QKD round has aborted, Eve could obtain the $j$-th bit value of the first QKD session key. Alternatively, Eve might also leak the key material produced in a certain QKD round by simply hiding it in the public discussion of subsequent QKD runs. We refer the reader to Ref.~\cite{mem_attack} for more details. Memory attacks are a fundamental threat to the security of DI-QKD.}
\label{mem_att}
\end{figure}
Despite its conceptual beauty, DI-QKD is however not foolproof. Indeed, one cannot expect that all QKD users will have expertise in experimental quantum optics and electronics. So, unless Alice and Bob manufacture their own QKD devices themselves, it could be very hard for them to guarantee that the devices are indeed honest. For instance, it was shown in~\cite{mem_attack} that DI-QKD is highly vulnerable to the so-called memory attacks. In this type of attacks, a hidden memory device (planted by the eavesdropper, Eve, in say Alice's setup during the manufacturing or initial installation of the QKD system) stores up the key material generated in each QKD session and then leaks this information to Eve in subsequent QKD runs. This situation is illustrated in Fig.~\ref{mem_att}. Importantly, such leakage of key information could be done very slowly over many subsequent QKD runs, and thus it could be very difficult to detect~\cite{mem_attack}. Obviously, this is a fatal loss of security for DI-QKD. Whenever a QKD system is reused for subsequent QKD sessions, the security of the keys generated in previous QKD runs might be compromised.~This is particularly problematic in a network setting with multiple users (who may not all be trustworthy) due to the impostor attack~\cite{mem_attack,counter_mem_attack}. Moreover, note that in principle memory attacks could also work for non-DI-QKD. This is so because, in practice, it could be quite challenging to check wether or not a purchased QKD setup contains such memory. Therefore, in the following, whenever we refer to a QKD system, it will be implicitly understood that it could be either a DI-QKD scheme or a non-DI-QKD scheme, as our results apply to both frameworks.

Our view is that memory attacks constitute an example of covert channels~\cite{covert}, which have attracted massive attention in conventional cryptography. With covert channels, seemingly innocent communications in a protocol could leak crucial information that is fatal to its security. One main motivation of our work is indeed to counter covert channels, such as memory attacks, in QKD.

\begin{figure}
  \includegraphics[width=0.95\columnwidth]{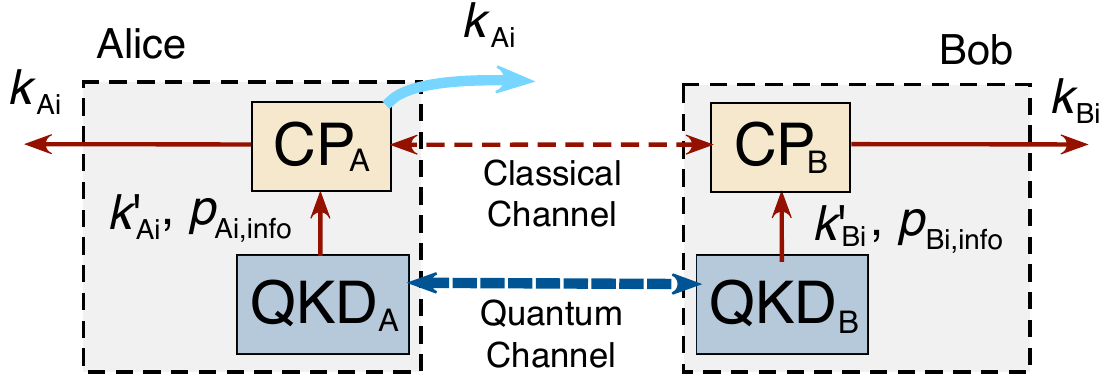}
\caption{Schematic representation of a hardware/software Trojan Horse attack against a QKD system. In QKD, it is commonly and implicitly assumed that the classical post-processing units CP$_{\rm A}$ and CP$_{\rm B}$ are trusted. However, due to the many hardware~\cite{covert2,covert2b,covert2c} and software~\cite{covert3} Trojan Horse attacks that have been performed against conventional cryptographic systems, this trust is a very strong and unjustified assumption. For instance, Eve could modify a chip, or infect the software with malware, to make it fail at a crucial time, or to hide a backdoor in say Alice's unit CP$_{\rm A}$ that leaks the final key, $k_{{\rm A}i}$, generated in say the $i$-th QKD run to her~\cite{military1,military2}. Due to the complexity and fabrication costs of current chips, these devices are typically designed by different parties, manufactured by an external foundry, and packaged and distributed by separate companies. This gives Eve multiple opportunities to meddle with the hardware. More importantly, hardware Trojans can be very hard to detect in practice. This is so because even slight adjustments to the electrical properties of a few transistors (from the billions of them contained in today's chips) could already compromise the security.~Also, Eve could easily bypass post-fabrication tests by crafting attack triggers that require a sequence of unlikely events, or by altering only a subset of the chips in question~\cite{covert2b,covert2c}. Similar arguments apply as well to software malware. This shows the weaknesses of the classical post-processing units. We refer the reader to the caption of Fig.~\ref{mem_att} for the meaning of the different elements in this figure.}
\label{fig0}
\end{figure}

Another key weakness in standard QKD security proofs is that they all implicitly assume that the classical post-processing units are trusted. These units are supposed to distill a secure secret key from the raw data generated by the QKD modules by applying techniques such as post-selection of data (or so-called sifting), parameter estimation, error correction, error verification and privacy amplification. However, in view of the many hardware~\cite{covert2,covert2b,covert2c} and software~\cite{covert3} Trojan Horse attacks that have been performed recently in conventional cryptographic systems, such trust is a very strong and unjustified assumption. This scenario is illustrated in Fig.~\ref{fig0}. Indeed, hardware and software Trojans constitute today a key threat to the security of conventional cryptographic devices and this threat is expected to only rise with time, so it cannot be neglected when analysing the security of a QKD implementation.

And so the key question is: how do we address covert channels and prove security in QKD with untrusted classical post-processing units? The existence of memory attacks in DI-QKD shows that quantum mechanics alone is not enough. Clearly, we need to include some additional assumptions. To solve this problem, we draw inspiration from the idea of verifiable secret sharing (VSS)~\cite{verifiable1,verifiable2} and the existence of secure multiparty computations~\cite{book_smc} in conventional cryptography, where it is known that one can achieve information-theoretic security in a $n$-party cryptographic setup if the number of cheaters is less than $n/3$~\cite{smc1,smc2,maurer_smpc}.

Standard VSS schemes, however, assume that all channels are classical, so by using error correction and authentication techniques one  can basically make these channels perfect.~In contrast, in QKD, owing to the noisy and lossy quantum channels controlled by Eve and the quantum no-cloning theorem, to distill a final key Alice and Bob need to apply several classical post-processing steps to the raw data produced by the QKD modules in a setting where {\it both} the QKD modules and the classical post-processing units might be corrupted.

\begin{figure}
  \includegraphics[width=0.96\columnwidth]{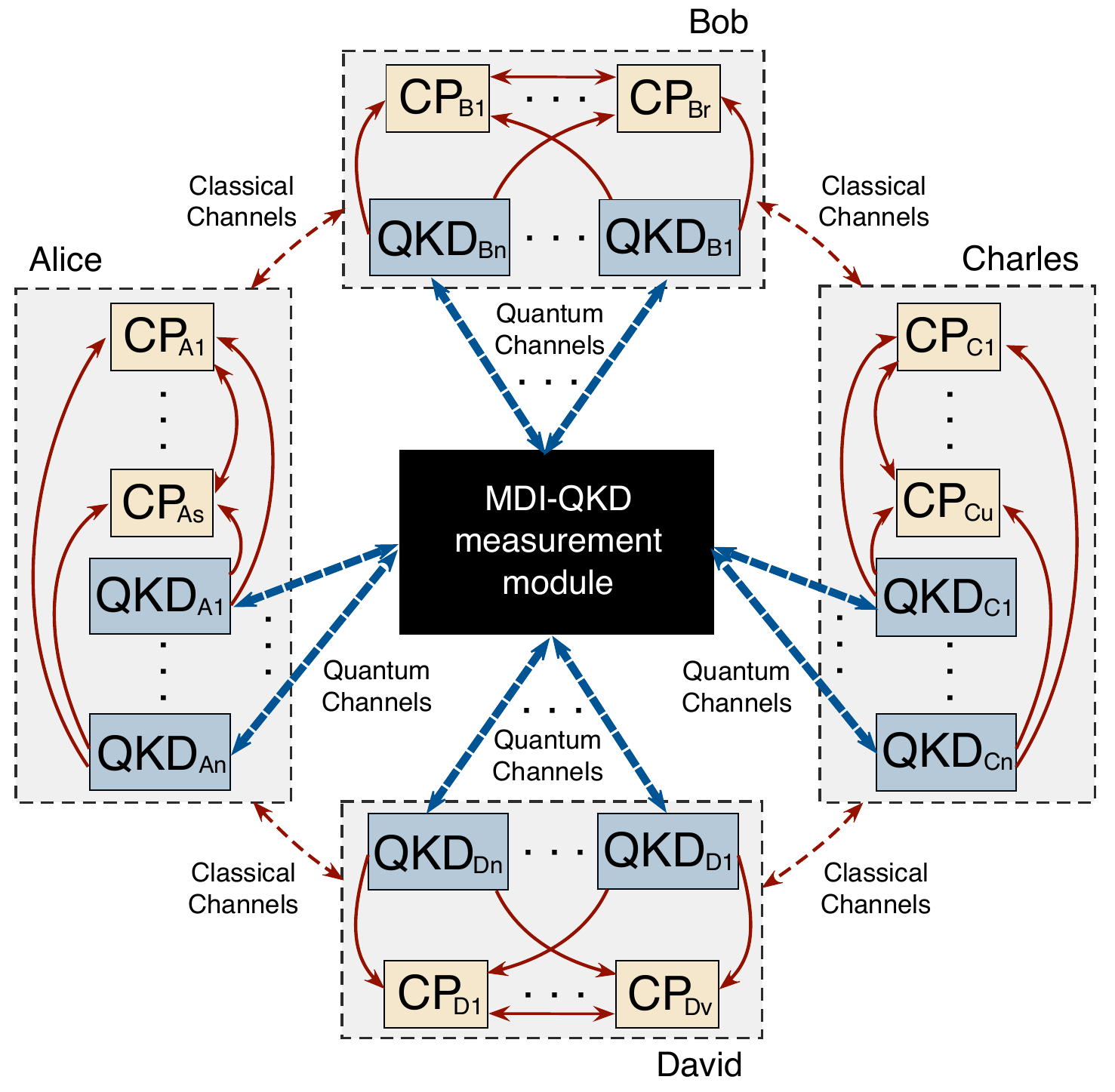}
\caption{Schematic representation of a MDI-QKD network~\cite{mdiQKD} in which each user has multiple QKD transmitter modules and classical post-processing units. Note that the measurement devices are often the most expensive components of an entire QKD system because single photon detection is highly non-trivial. MDI-QKD allows measurement modules to be totally untrusted, which means that there is no need for redundant measurement modules if our proposal is employed with MDI-QKD. The users just need to have multiple transmitters and classical post-processing units, which thanks to the development of cheap chip-based QKD systems~\cite{chip1,chip2,chip3}, we believe, could render our proposal cost effective in the future. We remark that our approach is also fully compatible with quantum relays and quantum repeaters.}
\label{fig_mdi}
\end{figure}
A key contribution of this paper is thus to show how, despite these obstacles, such VSS approach could be adapted to QKD to re-establish its security. The price that we pay is that now Alice and Bob have to use a redundant number of QKD modules and classical post-processing units. Fortunately, however, with the recent development of measurement-device-independent QKD (MDI-QKD)~\cite{mdiQKD,mdiQKDe1,mdiQKDe2,mdiQKDe3,mdiQKDe4} and chip-based QKD~\cite{chip1,chip2,chip3}, the cost of QKD modules might decrease dramatically over time, see Fig.~\ref{fig_mdi}. So, it is not unrealistic to consider that each of Alice and Bob could possess a few QKD modules and classical post-processing units, each of them purchased from a different vendor. Now, provided that the majority of the vendors are honest and careful in the manufacturing of their devices, it might not be entirely unreasonable to assume that at least one pair of QKD modules is honest and the number of malicious/flawed classical post-processing units is strictly less than one third of the total number of them. With these assumptions in place, we can then apply techniques in conventional multiparty secure computation to prove security in different QKD scenarios with malicious devices. Importantly, if we disregard the cost of authenticating the classical communications, our protocols are optimal with respect to the resulting secret key rate. Moreover, the operations involved are based on simple functions in linear algebra such as bit-wise XOR and multiplication of matrices.~So, they are conceptually simple and easy to implement.

\section{QKD with malicious devices}\label{mainresults}

\begin{figure}
  \includegraphics[width=1\columnwidth]{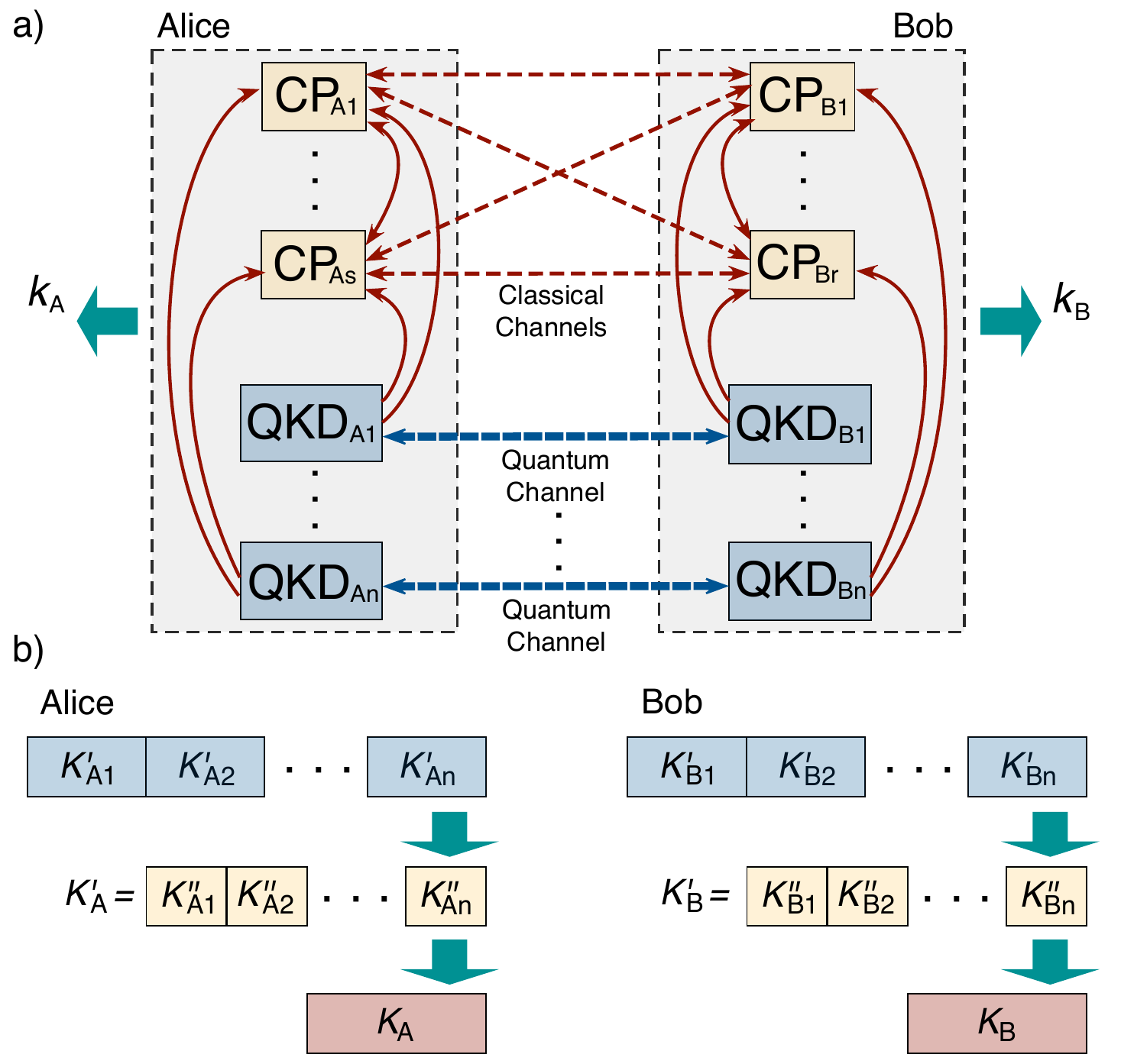}
\caption{{\bf (a)} Schematic representation of a QKD setup with multiple QKD modules and classical post-processing units. We assume that up to $t<n$ pairs of QKD modules, up to $t'<s/3$ units CP$_{{\rm A}i'}$ and  $t''<r/3$ units CP$_{{\rm B}i''}$ could be corrupted. The goal is to distill shares of an $\epsilon$-secure key, $k_{\rm A}$ and $k_{\rm B}$. In the figure, the thin red solid lines represent secure classical channels, the thin red dashed lines denote authenticated classical channels, and the blue thick dashed lines are quantum channels. {\bf (b)} General strategy to distill shares of $k_{\rm A}$ and $k_{\rm B}$. First, each pair QKD$_{{\rm A}i}$ and QKD$_{{\rm B}i}$ outputs a raw key, $k'_{{\rm A}i}$ and $k'_{{\rm B}i}$, together with the protocol information, and sends them to the CP units at Alice and Bob's side respectively. From $k'_{{\rm A}i}$ and $k'_{{\rm B}i}$, these units distill a supposedly $(\epsilon/n)$-secure key, $k''_{{\rm A}i}$ and $k''_{{\rm B}i}$, and then concatenate these keys to form $k'_{\rm A}=[k''_{{\rm A}1}, k''_{{\rm A}2}, \ldots, k''_{{\rm A}n}]$ and $k'_{\rm B}=[k''_{{\rm B}1}, k''_{{\rm B}2}, \ldots, k''_{{\rm B}n}]$. Finally, the CP units apply privacy amplification to $k'_{\rm A}$ and $k'_{\rm B}$ to remove the information held by the corrupted QKD modules and obtain $k_{\rm A}$ and $k_{\rm B}$. In the presence of corrupted CP units, all these steps are realised in a distributed setting by acting on data shares generated with a VSS scheme.
}
\label{fig_gen}
\end{figure}

Let us start by describing the general scenario that we consider in more detail. It is illustrated in Fig.~\ref{fig_gen}(a). Alice and Bob have $n$ pairs of QKD modules, and, in addition, say Alice (Bob) has $s$ ($r$) classical post-processing units at their disposal, each of them ideally purchased from a different provider. Alice's modules QKD$_{{\rm A}i}$, with $i=1,\ldots,n$, are connected to the classical post-processing units CP$_{{\rm A}i'}$, with $i'=1,\ldots,s$, via secure channels ({\it i.e.}, channels that provide both secrecy and authentication). Also, all the units CP$_{{\rm A}i'}$ are connected to each other via secure channels. The same applies to Bob. Importantly, since all these secure channels are located only within Alice and Bob's labs, in practice they could be implemented, for instance, by using physically protected paths ({\it e.g.}, physical wires that are mechanically and electrically protected against damage and intrusion) which connect only the prescribed devices. Furthermore, each QKD$_{{\rm A}i}$ is connected to its partner QKD$_{{\rm B}i}$ via a quantum channel, and each CP$_{{\rm A}i'}$ is connected to all CP$_{{\rm B}i''}$, with $i'=1,\ldots,s$ and $i''=1,\ldots,r$, via authenticated classical channels~\cite{wegman,wegman2}.

Moreover, for simplicity, we shall consider a so-called threshold active adversary structure. That is, we will assume that up to $t<n$ pairs of QKD modules, up to $t'<s/3$ units CP$_{{\rm A}i'}$ and up to $t''<r/3$ independent units CP$_{{\rm B}i''}$ could be corrupted. We say that a pair of QKD modules is corrupted when at least one of them is corrupted. Also, we conservatively assume that corrupted devices do not have to necessarily follow the prescriptions of the protocol but their behaviour is fully controlled by Eve, who could also access all their internal information. We refer the reader to Appendix~\ref{gen_sec} for the security analysis of QKD against a general mixed adversary structure~\cite{fitzi}.

The goal is to generate a composable $\epsilon$-secure key, $k_{\rm A}$ and $k_{\rm B}$. That is, $k_{\rm A}$ and $k_{\rm B}$ should be identical except for a minuscule probability $\epsilon_{\rm cor}$, and say $k_{\rm A}$ should be completely random and decoupled from Eve except for a minuscule probability $\epsilon_{\rm sec}$, with $\epsilon_{\rm cor}+\epsilon_{\rm sec}\leq\epsilon$~\cite{comp1,comp2}. Importantly, since now some QKD modules and classical post-processing units could be corrupted, the secrecy condition also implies that $k_{\rm A}$ and $k_{\rm B}$ must be independent of any information held by the corrupted devices {\it after} the execution of the protocol. Otherwise, such corrupted devices could directly leak $k_{\rm A}$ and $k_{\rm B}$ to Eve. Obviously, at the end of the day, some parties might need to have access to the final key, and thus one necessarily must assume that such parties are trusted and located in secure labs. In this regard, our work suggests that when the classical post-processing units at the key distillation layer are untrusted, they should not output the final key $k_{\rm A}$ and $k_{\rm B}$ but they should output shares of it to the key management layer~\cite{kmang1,kmang2}. There, $k_{\rm A}$ and $k_{\rm B}$ could be either reconstructed by say Alice and Bob in secure labs, or its shares could be stored in distributed memories for later use, or they could be employed for instance for encryption purposes via say the one-time pad. Importantly, however, all the key generation process at the key distillation layer can be performed with corrupted devices. Also, we note that, if necessary, operations like storage or encryption at the key management layer could also be performed with corrupted devices by using techniques from secure multiparty computation~\cite{book_smc}. In any case, the actual management and storage of the shares of $k_{\rm A}$ and $k_{\rm B}$ generated by the key distillation layer is responsibility of the key management layer and depends on the particular application.

Before we address specific scenarios in detail, let us provide an overview of the general strategy that we follow to achieve our goal, which uses as main ingredients VSS schemes~\cite{verifiable1,verifiable2,maurer_smpc} and privacy amplification techniques (see Appendix~\ref{tool}). The former is employed to defeat corrupted classical post-processing units. Indeed, given that $t'<s/3$ and $t''<r/3$, the use of VSS schemes allows to post-process the raw keys generated by the QKD modules in a distributed setting by acting only on raw key shares. More importantly, this post-processing of raw key shares can be performed such that no set of corrupted classical post-processing units can reconstruct $k_{\rm A}$ and $k_{\rm B}$ and, moreover, it is also guaranteed that $k_{\rm A}$ and $k_{\rm B}$ is a correct key independently of the misbehaviour of the corrupted units which might wish to purposely introduce errors. In this regard, a key insight of our paper is to show that, since all the classical post-processing techniques that are typically applied in QKD are ``linear'' in nature ({\it i.e.}, they involve simple functions in linear algebra such as bit-wise XOR and multiplications of matrices), they are easily implementable in a distributed setting.  

Let us illustrate this point with a simple example. In particular, let us consider, for instance, the error correction step in QKD. Here, say Bob wants to correct a certain bit string, $k_{\rm B, key}$, to match that of Alice, which we shall denote by $k_{\rm A, key}$. In general, this process requires that both Alice and Bob first apply certain error correction matrices, $M_{\rm EC}$, to $k_{\rm B, key}$ and $k_{\rm A, key}$ to obtain the syndrome information $s_{\rm A}=M_{\rm EC}k_{\rm A, key}$ and $s_{\rm B}=M_{\rm EC}k_{\rm B, key}$, respectively. Afterward, if $s_{\rm A}\neq{}s_{\rm B}$ Bob modifies $k_{\rm B, key}$ accordingly. This process is then repeated a few times until it is guaranteed that $k_{\rm B, key}=k_{\rm A, key}$ with high probability. Let us now consider again the same procedure but now acting on shares, $k_{\rm A{\it j}, key}$ and $k_{\rm B{\it j}, key}$, of $k_{\rm A, key}$ and $k_{\rm B, key}$ respectively. That is, say $k_{\rm A, key}=\oplus_j^q k_{\rm A{\it j}, key}$ and $k_{\rm B, key}=\oplus_j^q k_{\rm B{\it j}, key}$, with $q$ being the total number of shares. For this, Alice and Bob first apply $M_{\rm EC}$ to $k_{\rm A{\it j}, key}$ and $k_{\rm B{\it j}, key}$ to obtain $s_{{\rm A}j}=M_{\rm EC}k_{\rm A{\it j}, key}$ and $s_{{\rm B}j}=M_{\rm EC}k_{\rm B{\it j}, key}$, respectively, for all $j$. Next, Alice sends  $s_{{\rm A}j}$ to Bob who obtains $s_{\rm A}=\oplus_{j=1}^qs_{{\rm A}j}$ and $s_{\rm B}=\oplus_{j=1}^qs_{{\rm B}j}$. This is so because $\oplus_{j=1}^qs_{{\rm A}j}=\oplus_{j=1}^qM_{\rm EC}k_{\rm A{\it j}, key}=M_{\rm EC}\oplus_{j=1}^qk_{\rm A{\it j}, key}=M_{\rm EC}k_{\rm A, key}=s_{\rm A}$, and a similar argument applies to $s_{\rm B}$. Finally, if $s_{\rm A}\neq{}s_{\rm B}$ Bob corrects $k_{\rm B, key}$ by acting on its shares $k_{\rm B{\it j}, key}$. This is so because to flip certain bits in $k_{\rm B, key}$ is equivalent to flip the corresponding bits in one of its shares $k_{\rm B{\it j}, key}$. That is, the error correction step in QKD can be easily performed in a distributed setting by acting only on shares of $k_{\rm A, key}$ and $k_{\rm B, key}$. The same argument applies as well to the other classical post-processing techniques in QKD, as all of them involve only linear operations.

To defeat corrupted QKD modules, on the other hand, we use privacy amplification techniques. Suppose, for instance, that each pair of QKD modules, QKD$_{{\rm A}i}$ and  QKD$_{{\rm B}i}$, output a raw key, $k'_{{\rm A}i}$ and $k'_{{\rm B}i}$. Moreover, suppose for the moment that the classical post-processing units are trusted and they distill a supposedly $(\epsilon/n)$-secure key, $k''_{{\rm A}i}$ and $k''_{{\rm B}i}$, of length $N$ bits from each pair $k'_{{\rm A}i}$ and $k'_{{\rm B}i}$. Then, we have that the $n\times{}N$ bit strings $k'_{\rm A}=[k''_{{\rm A}1},\ldots, k''_{{\rm A}n}]$ and $k'_{\rm B}=[k''_{{\rm B}1},\ldots, k''_{{\rm B}n}]$ are for certain $\epsilon_{\rm cor}$-correct. The secrecy condition, however, only holds if all the QKD modules are trusted. If say the pair QKD$_{{\rm A}i'}$ and  QKD$_{{\rm B}i'}$ is corrupted then the key strings $k''_{{\rm A}i'}$ and $k''_{{\rm B}i'}$ are compromised. So, given that $t<n$ the classical post-processing units can apply privacy amplification to $k'_{\rm A}$ and $k'_{\rm B}$ to extract two shorter $(n-t)\times{}N$ bit strings, $k_{\rm A}$ and $k_{\rm B}$, which are for certain $\epsilon_{\rm sec}$-secret and thus $\epsilon$-secure. In the presence of untrusted classical post-processing units, this process can be performed in a distributed manner by acting on data shares, just as we describe above.

In short, the general strategy can be decomposed in three main steps, which are illustrated in Fig.~\ref{fig_gen}(b). First, each pair of QKD modules generates a raw key and the protocol information and sends them to the CP units. Then, in a second step, the CP units distill a supposedly $(\epsilon/n)$-secure key from each raw key received and concatenate the resulting keys to form a longer key bit string. Finally, in the third step, the CP units apply privacy amplification to remove the information that could be known to Eve due to the presence of corrupted QKD modules. Importantly, if the CP units are untrusted, all these steps are performed in a distributed setting by acting on data shares produced by a VSS scheme.

Next we evaluate three different scenarios of practical interest in this context. For concreteness, in these examples we use the VSS scheme introduced in~\cite{maurer_smpc} and described in Appendix~\ref{tool}. 

\subsection{QKD with malicious QKD modules and honest classical post-processing units}\label{ups} 

\begin{figure}
  \includegraphics[width=0.95\columnwidth]{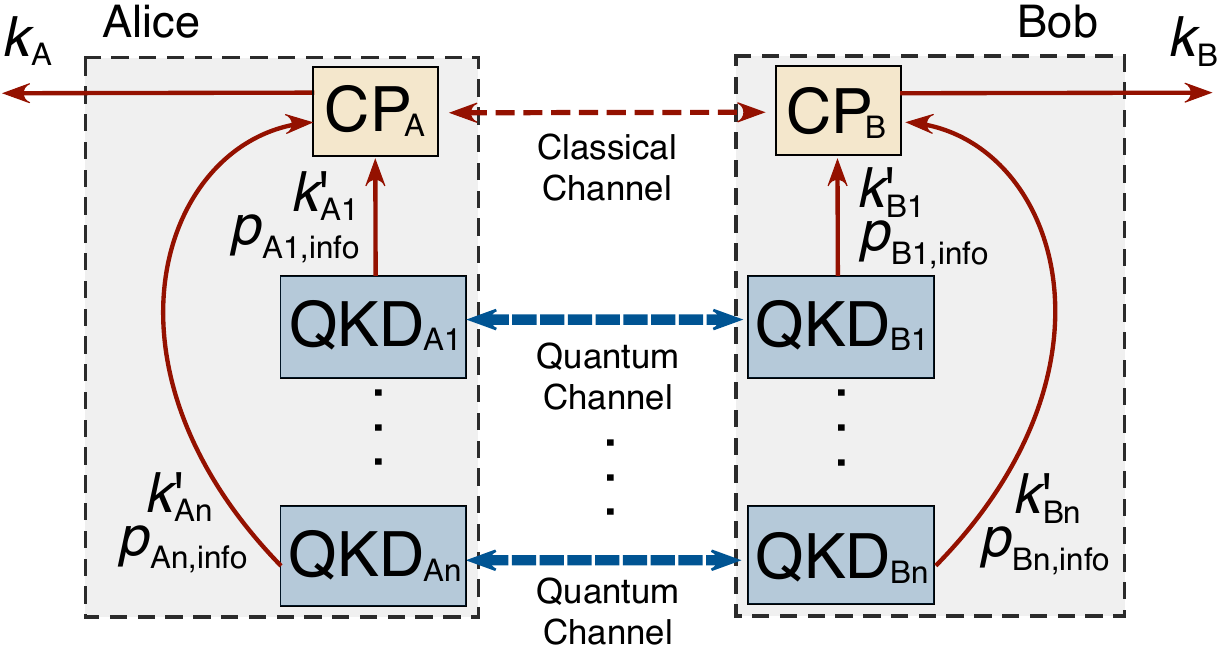}
\caption{QKD with malicious QKD modules and honest classical post-processing units. Alice and Bob have $n$ pairs of QKD modules, and up to $t<n$ of them could be corrupted. Alice's (Bob's) $i$-th QKD module is supposed to generate a raw key $k'_{{\rm A}i}$ ($k'_{{\rm B}i}$) and the protocol information $p_{\rm A{\it i},info}$ ($p_{\rm B{\it i},info}$), with $i=1,\ldots,n$. Also, they have one classical post-processing unit each, which is assumed to be honest. The goal is to distill an $\epsilon$-secure key, $k_{\rm A}$ and $k_{\rm B}$. This can be achieved by using {\it Protocol}~$1$. See the main text for further details.}
\label{fig3}
\end{figure}

We begin by analysing the situation where Alice and Bob have $n$ pairs of QKD modules and up to $t<n$ of them could be corrupted, and each of Alice and Bob has one classical post-processing unit which is assumed to be honest. This scenario is illustrated in Fig.~\ref{fig3} and corresponds to the case $s=r=1$ and $t'=t''=0$ in Fig.~\ref{fig_gen}(a).

A possible solution to this scenario is rather simple. It is given by {\it Protocol}~$1$ below, which consists of three main steps.
\\

\noindent{\it Protocol $1$}: 
\begin{enumerate}
\item {\it Generation of raw keys and protocol information}: Each pair QKD$_{{\rm A}i}$ and QKD$_{{\rm B}i}$ outputs, respectively, the bit strings $k'_{{\rm A}i}$ and $p_{\rm A{\it i},info}$, and $k'_{{\rm B}i}$ and $p_{\rm B{\it i},info}$, or the symbol $\perp_i$ to indicate abort, for all $i=1,\ldots,n$.
\item {\it Generation of an $\epsilon_{\rm cor}$-correct key}: The units CP$_{\rm A}$ and CP$_{\rm B}$ use the key distillation procedure prescribed by the QKD protocol to generate an $(\epsilon/n)$-secure key, $k''_{{\rm A}i}$ and $k''_{{\rm B}i}$, from each raw key pair $k'_{{\rm A}i}$ and $k'_{{\rm B}i}$, or they generate the abort symbol $\perp_i$, for all $i=1,\ldots, n$. Afterward, CP$_{\rm A}$ (CP$_{\rm B}$) concatenates the $M\leq{}n$ keys $k''_{{\rm A}i}$ ($k''_{{\rm B}i}$) which are different from $\perp_i$ to form the bit string $k'_{\rm A}=[k''_{{\rm A}1},\ldots, k''_{{\rm A}M}]$ ($k'_{\rm B}=[k''_{{\rm B}1},\ldots, k''_{{\rm B}M}]$).~Since the units CP$_{\rm A}$ and CP$_{\rm B}$ are trusted, $k''_{{\rm A}i}$ and $k''_{{\rm B}i}$ are for certain $(\epsilon_{\rm cor}/n)$-correct $\forall i$ and thus $k'_{\rm A}$ and $k'_{\rm B}$ are $\epsilon_{\rm cor}$-correct. The secrecy condition only holds if all $k''_{{\rm A}i}$ and $k''_{{\rm B}i}$ originate from raw keys output by honest QKD modules. For simplicity, we will suppose that the length of $k''_{{\rm A}i}$ and $k''_{{\rm B}i}$ is $N$ bits $\forall i$.
\item {\it Generation of an $\epsilon$-secure key}: CP$_{\rm A}$ and CP$_{\rm B}$ apply a randomly selected universal$_2$ hash function to extract from $k'_{\rm A}$ and $k'_{\rm B}$ two shorter bit strings, $k_{\rm A}$ and $k_{\rm B}$, of length $(M-t)\times{}N$ bits. $k_{\rm A}$ and $k_{\rm B}$ are by definition $\epsilon_{\rm sec}$-secret, and thus, from step~$2$, they are $\epsilon$-secure.
\end{enumerate}

Note that in step~$3$ of {\it Protocol}~$1$ we consider the worst-case scenario where all $k''_{{\rm A}i}$ and $k''_{{\rm B}i}$ generated by corrupted QKD modules contribute to $k'_{\rm A}$ and $k'_{\rm B}$ respectively, as Alice and Bob cannot discard this case. Most importantly, {\it Protocol}~$1$ allows Alice and Bob to defeat covert channels such as memory attacks in QKD, as this protocol guarantees that none of the corrupted QKD modules can access $k_{\rm A}$ or $k_{\rm B}$. Our results are summarised in the following Claim, whose proof is direct from the definition of {\it Protocol}~$1$.
\\

\noindent{\bf Claim 1.} {\it Suppose that Alice and Bob have $n$ pairs of QKD modules and up to $t<n$ of them could be corrupted. Also, suppose that they have one trusted classical post-processing unit each. Let $M\leq{}n$ denote the number of pairs of QKD modules that do not abort and whose raw key could in principle be transformed into an $(\epsilon/n)$-secure key, and let $N$ bits be the length of such supposedly secure key. Protocol~$1$ allows Alice and Bob to distill an $\epsilon$-secure key of length $(M-t)\times{}N$ bits. Moreover, the re-use of the devices does not compromise the security of the keys distilled in previous QKD runs.
}
\\

Importantly, we remark that {\it Protocol}~$1$ is optimal with respect to the resulting secret key rate. This is so because of the following. If no pair of QKD modules aborts and its raw data could in principle be transformed into a secure key we have, by definition, that the maximum {\it total} final key length is at most $n\times{}N$ bits. Also, we know that up to $t\times{}N$ bits of such key could be compromised by the $t$ pairs of corrupted QKD modules. That is, the maximum secure key length is at most $(n-t)\times{}N$ bits. Moreover, as discussed above, if some pairs of QKD modules abort we must necessarily assume the worst-case scenario where they are honest. This is so because through her interaction with the quantum signals in the channel, Eve could always force honest QKD modules to abort by simply increasing the resulting QBER or phase error rate. That is, in the scenario considered it is not possible to distill a key length greater than $(M-t)\times{}N$ bits.

\subsection{QKD with honest QKD modules and malicious classical post-processing units}\label{ups2} 

\begin{figure}
  \includegraphics[width=0.95\columnwidth]{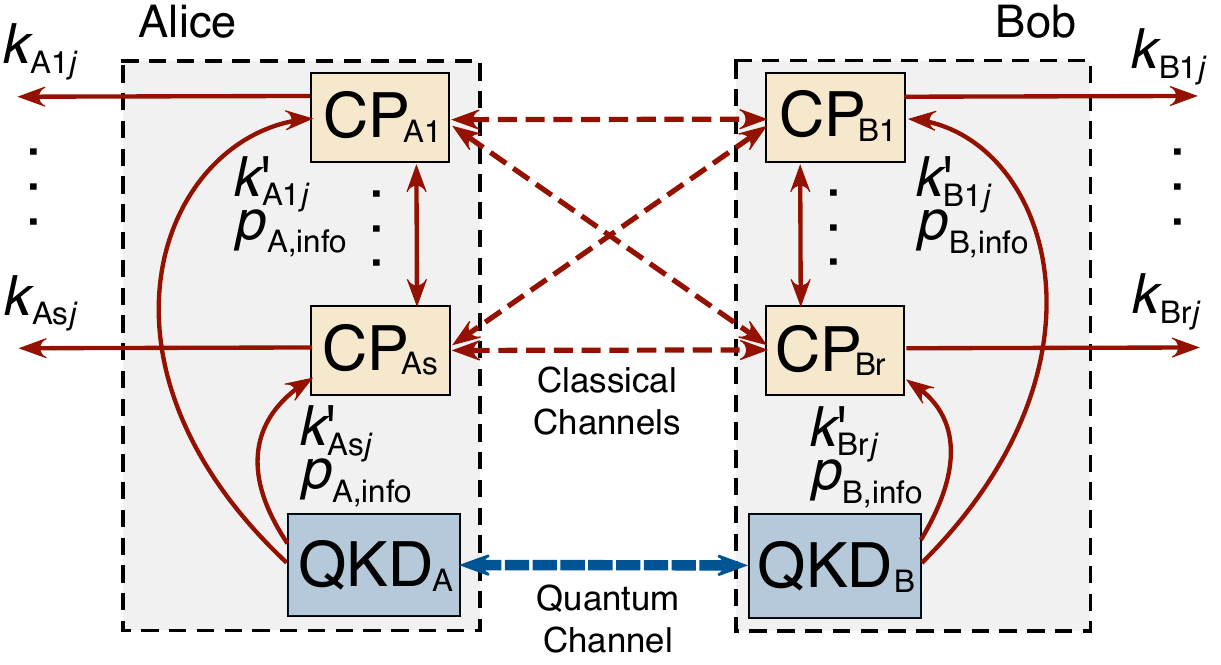}
\caption{QKD with honest QKD modules and malicious classical post-processing units. Alice and Bob have one trusted QKD module each, QKD$_{\rm A}$ and QKD$_{\rm B}$, which generate, respectively, the raw key $k'_{\rm A}$ and $k'_{\rm B}$ and the protocol information $p_{\rm A,info}$ and $p_{\rm B,info}$. Also, Alice (Bob) has $s$ ($r$) classical post-processing units CP$_{{\rm A}i}$ (CP$_{{\rm B}i'}$) with $i=1,\ldots,s$ ($i'=1,\ldots,r$), and up to $t'<s/3$ ($t''<r/3$) of these units could be corrupted. The goal is to produce shares of an $\epsilon$-secure key, $k_{\rm A}$ and $k_{\rm B}$, from which the final key could be reconstructed. This can be achieved by using {\it Protocol}~$2$. See the main text and Appendix~\ref{app_P2} for further details. In the figure, $k'_{{\rm A}ij}$ ($k'_{{\rm B}i'j}$) denotes the $j$-th share of $k'_{\rm A}$ ($k'_{\rm B}$) which QKD$_{\rm A}$ (QKD$_{\rm B}$) sends to CP$_{{\rm A}i}$ (CP$_{{\rm B}i'}$), and $k_{{\rm A}ij}$ ($k_{{\rm B}ij}$) identifies the $j$-th share of $k_{\rm A}$ ($k_{\rm B}$) that is produced by CP$_{{\rm A}i}$ (CP$_{{\rm B}i'}$). Since QKD$_{\rm A}$ (QKD$_{\rm B}$) is honest, note that the shares $k'_{{\rm A}ij}$ ($k'_{{\rm B}ij}$) are equal for all $i$.}
\label{fig2}
\end{figure}

We now consider the situation where Alice and Bob have one trusted QKD module each, and Alice (Bob) has $s$ ($r$) classical post-processing units CP$_{{\rm A}i}$ (CP$_{{\rm B}i'}$), with $i=1,\ldots,s$ ($i'=1,\ldots,r$), and up to $t'<s/3$ ($t''<r/3$) of them could be corrupted. This scenario is illustrated in Fig.~\ref{fig2} and corresponds to the case $n=1$ and $t=0$ in Fig.~\ref{fig_gen}(a).

Since now the units CP$_{{\rm A}i}$ and CP$_{{\rm B}i'}$ could be malicious, we aim to generate shares of an $\epsilon$-secure key, $k_{\rm A}$ and $k_{\rm B}$. A possible solution to this scenario 
is given by {\it Protocol}~$2$, which uses the VSS protocol introduced in~\cite{maurer_smpc}. These protocols are described in Appendices~\ref{tool} and \ref{app_P2} respectively. Below we provide a sketch of {\it Protocol}~$2$ where, for easy of presentation, we assume $r=s$. It consists of six main steps.
\\

\noindent Sketch of {\it Protocol $2$}: 
\begin{enumerate}
\item {\it Generation and distribution of shares of raw keys and protocol information}: QKD$_{\rm A}$ and QKD$_{\rm B}$ output, respectively, $k'_{\rm A}$ and $p_{\rm A,info}$, and $k'_{\rm B}$ and $p_{\rm B,info}$, or the abort symbol $\perp$. If the output is different from $\perp$, QKD$_{\rm A}$ sends $p_{\rm A,info}$ to all CP$_{{\rm A}i}$ and uses a VSS scheme to distribute shares of $k'_{\rm A}$ between the CP$_{{\rm A}i}$. Likewise, QKD$_{\rm B}$ does the same with $p_{\rm B,info}$, $k'_{\rm B}$ and the units CP$_{{\rm B}i'}$. Let $k'_{{\rm A}ij}$ ($k'_{{\rm B}i'j}$) be the $j$-th share of $k'_{\rm A}$ ($k'_{\rm B}$) received by CP$_{{\rm A}i}$ (CP$_{{\rm B}i'}$). Next, the CP$_{{\rm A}i}$ and CP$_{{\rm B}i'}$ send to each other $p_{\rm A,info}$ and $p_{\rm B,info}$.
\item {\it Sifting}: Each CP$_{{\rm A}i}$ uses $p_{\rm A,info}$ and $p_{\rm B,info}$ to obtain two bit strings, $k_{\rm A{\it ij}, key}$ and $k_{\rm A{\it ij}, est}$, from $k'_{{\rm A}ij}$. $k_{\rm A{\it ij}, key}$ ($k_{\rm A{\it ij}, est}$) is used for key generation (parameter estimation). Likewise, Bob's CP$_{{\rm B}i'}$ do the same with $k'_{{\rm B}i'j}$.
\item {\it Parameter estimation}: The CP$_{{\rm A}i}$ and CP$_{{\rm B}i'}$ use the reconstruct protocol of a VSS scheme to obtain the parts of $k'_{\rm A}$ and $k'_{\rm B}$ that are used for parameter estimation, $k_{\rm A, est}$ and $k_{\rm B, est}$, from their shares $k_{\rm A{\it ij}, est}$ and $k_{\rm B{\it i'j}, est}$. Next, each CP$_{{\rm A}i}$ and CP$_{{\rm B}i'}$ performs locally parameter estimation ({\it e.g.}, they estimate the phase error rate). If the estimated values exceed certain tolerated values, they abort. 
\item {\it Error correction}: The CP$_{{\rm A}i}$ and CP$_{{\rm B}i'}$ correct the parts of $k'_{\rm A}$ and $k'_{\rm B}$ that are used for key distillation, $k_{\rm A, key}$ and $k_{\rm B, key}$, by acting on their shares $k_{\rm A{\it ij}, key}$ and $k_{\rm B{\it i'j}, key}$. Let ${\hat k}_{\rm A{\it ij}, key}$ and ${\hat k}_{\rm B{\it i'j}, key}$ denote the resulting shares of the corrected keys ${\hat k}_{\rm A, key}$ and ${\hat k}_{\rm B, key}$, and let leak$_{\rm EC}$ bits be the syndrome information interchanged during this step. 
\item {\it Error verification}: The CP$_{{\rm A}i}$ use the RBS scheme (see Appendix~\ref{tool}) to randomly select a universal$_2$ hash function, $h_{\rm V}$, that is sent to all CP$_{{\rm B}i'}$. The CP$_{{\rm A}i}$ (CP$_{{\rm B}i'}$) compute $h_{{\rm A}ij}=h_{\rm V}({\hat k}_{\rm A{\it ij}, key})$ ($h_{{\rm B}i'j}=h_{\rm V}({\hat k}_{\rm B{\it i'j}, key})$) of length $\lceil\log_2{(4/\epsilon_{\rm cor})}\rceil$ bits, and they use the reconstruct protocol of a VSS scheme to obtain both a hash, $h_{\rm A}$, of ${\hat k}_{\rm A, key}$ and a hash, $h_{\rm B}$, of ${\hat k}_{\rm B, key}$ from the shares $h_{{\rm A}ij}$ and $h_{{\rm B}i'j}$. Finally, each CP$_{{\rm A}i}$ and CP$_{{\rm B}i'}$ aborts if $h_{\rm A}\neq{}h_{\rm B}$. 
\item {\it Generation of shares of an $\epsilon$-secure key}: The CP$_{{\rm A}i}$ use the RBS scheme to randomly select a universal$_2$ hash function, $h_{\rm P}$, that is sent to all CP$_{{\rm B}i'}$. Each CP$_{{\rm A}i}$ (CP$_{{\rm B}i'}$) obtains the shares, $k_{{\rm A}ij}=h_{\rm P}({\hat k}_{\rm A{\it ij}, key})$ ($k_{{\rm B}i'j}=h_{\rm P}({\hat k}_{\rm B{\it i'j}, key})$), of a key $k_{\rm A}$ ($k_{\rm B}$).
\end{enumerate}

Given that $t'<M_{\rm A}/3$ and $t''<M_{\rm B}/3$, where $M_{\rm A}$ ($M_{\rm B}$) denotes the number of CP$_{{\rm A}i}$ (CP$_{{\rm B}i'}$) that do not abort, we have that the final key, $k_{\rm A}$ and $k_{\rm B}$, is $\epsilon$-secure (see Appendix~\ref{app_P2}). If they wish so, Alice (Bob) can obtain $k_{\rm A}$ ($k_{\rm B}$) by using the reconstruct protocol of a VSS scheme. That is, Alice (Bob) can use majority voting to obtain first the $j$-th share, $k_{{\rm A}j}$ ($k_{{\rm B}j}$), of $k_{\rm A}$ ($k_{\rm B}$) from $k_{{\rm A}ij}$ ($k_{{\rm B}i'j}$) for all $j=1,\ldots,q$, and then she (he) calculates $k_{\rm A}=\oplus_{j=1}^q k_{{\rm A}j}$ ($k_{\rm B}=\oplus_{j=1}^q k_{{\rm B}j}$), where $q$ is the total number of shares.

Our results are summarised in the following Claim, whose proof is direct from the definition of {\it Protocol}~$2$:
\\

\noindent{\bf Claim 2.} {\it Suppose that Alice and Bob have one trusted QKD module each, and each of them has, respectively, $s$ and $r$ classical post-processing units. Also, suppose that up to $t'<s/3$ of Alice's units and up to $t''<r/3$ of Bob's units could be corrupted. Then, if we disregard the cost of authenticating the classical channels between Alice and Bob's classical post-processing units, Protocol~$2$ allows them to distill an $\epsilon$-secure key of the same length as would be possible in a completely trusted scenario. Moreover, the re-use of the devices does not compromise the security of the keys distilled in previous QKD runs.
}
\\

We remark that if we ignore the cost of authenticating the classical channels between the units CP$_{{\rm A}i}$ and CP$_{{\rm B}i'}$, Claim~$2$ implies directly that {\it Protocol}~$2$ is optimal with respect to the resulting secret key length. Also, we refer the reader to Appendix~\ref{alter} for a simpler but less efficient protocol to achieve the same task.

\subsection{QKD with malicious QKD modules and malicious classical post-processing units}\label{ups3}

\begin{figure}
  \includegraphics[width=0.95\columnwidth]{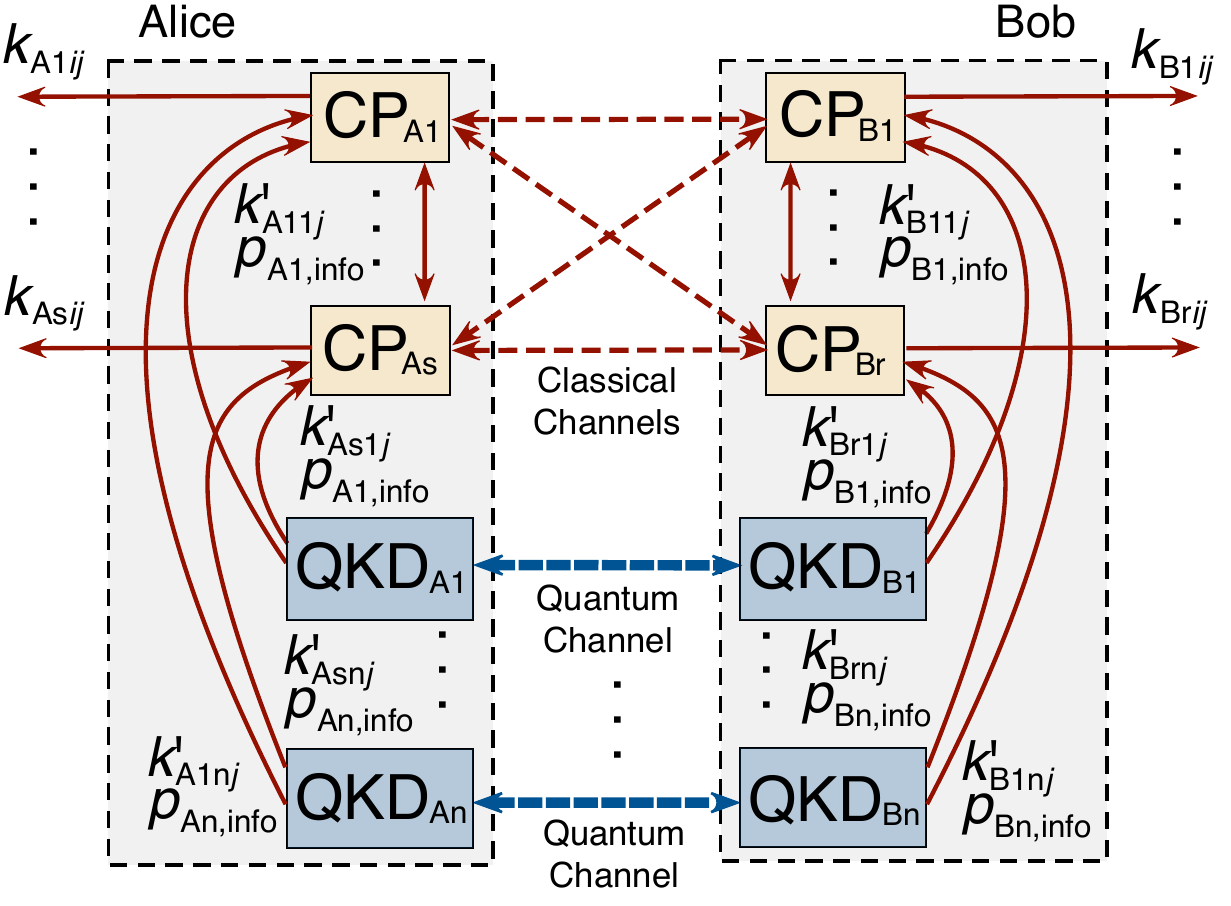}
\caption{QKD with malicious QKD modules and malicious classical post-processing units. Alice and Bob have $n$ pairs of QKD modules, QKD$_{{\rm A}i}$ and QKD$_{{\rm B}i}$, and up to $t<n$ of them could be corrupted. They generate, respectively, the raw key $k'_{{\rm A}i}$ and $k'_{{\rm B}i}$ and the protocol information $p_{\rm A{\it i},info}$ and $p_{\rm B{\it i},info}$. Also, Alice (Bob) has $s$ ($r$) classical post-processing units CP$_{{\rm A}i'}$ (CP$_{{\rm B}i''}$) with $i'=1,\ldots,s$ ($i''=1,\ldots,r$), and up to $t'<s/3$ ($t''<r/3$) of these units could be corrupted. The goal is to produce shares of an $\epsilon$-secure key, $k_{\rm A}$ and $k_{\rm B}$, from which the final key could be reconstructed. This can be achieved by using {\it Protocol}~$3$. See the main text for further details. In the figure, $k'_{{\rm A}i'ij}$ ($k'_{{\rm B}i''ij}$) denotes the $j$-th share of $k'_{{\rm A}i}$ ($k'_{{\rm B}i}$) which QKD$_{{\rm A}i}$ (QKD$_{{\rm B}i}$) sends to CP$_{{\rm A}i'}$ (CP$_{{\rm B}i''}$), and $k_{{\rm A}i'ij}$ ($k_{{\rm B}i''ij}$) identifies the shares of $k_{\rm A}$ ($k_{\rm B}$) that are produced by CP$_{{\rm A}i'}$ (CP$_{{\rm B}i''}$).
 }
\label{fig1}
\end{figure}

Finally, here we consider the situation where Alice and Bob have $n$ pairs of QKD modules, QKD$_{{\rm A}i}$ and QKD$_{{\rm B}i}$ with $i=1,\ldots,n$, and Alice (Bob) has $s$ ($r$) classical post-processing units CP$_{{\rm A}i'}$ (CP$_{{\rm B}i''}$), with $i'=1,\ldots,s$ ($i''=1,\ldots,r$), and up to $t<n$ pairs of QKD modules, up to $t'<s/3$ units CP$_{{\rm A}i'}$ and up to $t''<r/3$ units CP$_{{\rm B}i''}$ could be corrupted. This scenario is illustrated in Fig.~\ref{fig1} and corresponds to the most general case considered in Fig.~\ref{fig_gen}(a).

For illustrative purposes, let us discuss first a naive protocol that fails to achieve the goal. In particular, suppose for simplicity that $s=r=n$, and, moreover, we have that up to $t<n$ groups $G_i\equiv\{{\rm QKD}_{{\rm A}i}, {\rm QKD}_{{\rm B}i}, {\rm CP}_{{\rm A}i}, {\rm CP}_{{\rm B}i}\}$ could be corrupted, where we say that a group $G_i$ is corrupted if at least one of its elements is corrupted. Then, if one disregards efficiency issues, a straightforward solution to this scenario might appear to be as follows. Each $G_i$ simply generates a supposedly $\epsilon$-secure key, $k_{{\rm A}i}$ and $k_{{\rm B}i}$, and this key is then considered as the $i$-th share of a final key, $k_{\rm A}$ and $k_{\rm B}$. That is, $k_{\rm A}=\oplus_{i=1}^n k_{{\rm A}i}$ and $k_{\rm B}=\oplus_{i=1}^n k_{{\rm B}i}$. Indeed, given that $t<n$, $k_{\rm A}$ and $k_{\rm B}$ is for certain $\epsilon_{\rm sec}$-secret. However, the main problem of this naive approach is that $k_{\rm A}$ and $k_{\rm B}$ might not be {\it correct} because a corrupted $G_i$ could simple output $k_{{\rm A}i}\neq{}k_{{\rm B}i}$ and thus $k_{\rm A}\neq{}k_{\rm B}$. 

Below we provide a simple solution ({\it Protocol~$3$}) to the general scenario. It builds on {\it Protocols}~$1$ and $2$ above, and it consists of three main steps.
\\

\noindent{\it Protocol $3$}: 
\begin{enumerate}
\item {\it Generation and distribution of shares of $(\epsilon/n)$-secure keys}: Each pair QKD$_{{\rm A}i}$ and QKD$_{{\rm B}i}$ uses say {\it Protocol}~$2$ to distribute shares of an $(\epsilon/n)$-secure key, $k_{{\rm A}i}$ and $k_{{\rm B}i}$, or the abort symbol $\perp_i$, between CP$_{{\rm A}i'}$ and CP$_{{\rm B}i''}$, respectively. Let ${\tilde k}_{{\rm A}i'ij}$ (${\tilde k}_{{\rm B}i''ij'}$) be the $j$-th ($j'$-th) share of $k_{{\rm A}i}$ ($k_{{\rm B}i}$) obtained by CP$_{{\rm A}i'}$ (CP$_{{\rm B}i''}$). For simplicity, we will suppose that the length of $k_{{\rm A}i}$  and $k_{{\rm B}i}$ is $N$ bits for all $i$. 
\item {\it Generation of shares of an $\epsilon_{\rm cor}$-correct key}: Let $\vec{0}$ be the $N$-bit zero vector, and $M$ be the number of $k_{{\rm A}i}$ and $k_{{\rm B}i}$ which are different from $\perp_i$. Each CP$_{{\rm A}i'}$ defines $k''_{{\rm A}i'ij}=[\vec{0}_1,\ldots, \vec{0}_{i-1},{\tilde k}_{{\rm A}i'ij},\vec{0}_{i+1},\ldots, \vec{0}_{M}]$. Likewise, the CP$_{{\rm B}i''}$ form $k''_{{\rm B}i''ij'}$ from ${\tilde k}_{{\rm B}i''ij'}$. $k''_{{\rm A}i'ij}$ and $k''_{{\rm B}i''ij'}$ are by definition shares of an $\epsilon_{\rm cor}$-correct key. The secrecy condition only holds if all $k_{{\rm A}i}$  and $k_{{\rm B}i}$ originate from honest QKD modules.
\item {\it Generation of shares of an $\epsilon$-secure key}: The CP$_{{\rm A}i'}$ use the RBS scheme (see Appendix~\ref{tool}) to randomly select a universal$_2$ hash function, $h_{\rm P}$, that is sent to all CP$_{{\rm B}i''}$. Each CP$_{{\rm A}i'}$ (CP$_{{\rm B}i''}$) obtains shares, $k_{{\rm A}i'ij}=h_{\rm P}(k''_{{\rm A}i'ij})$ ($k_{{\rm B}i''ij'}=h_{\rm P}(k''_{{\rm B}i''ij'})$), of length $(M-t)\times{}N$ bits of a final key $k_{\rm A}$ ($k_{\rm B}$).
\end{enumerate}

Indeed, given that $t'<M_{{\rm A}i}/3$ and $t''<M_{{\rm B}i}/3$ for all $i=1,\ldots,M$, where $M_{{\rm A}i}$ ($M_{{\rm B}i}$) denotes the number of CP$_{{\rm A}i'}$ (CP$_{{\rm B}i''}$) that do not produce $\perp_i$ but output post-processed shares, $k_{{\rm A}i'ij}$ ($k_{{\rm B}i''ij'}$), from $k_{{\rm A}i}$ ($k_{{\rm B}i}$), then the final key, $k_{\rm A}$ and $k_{\rm B}$, is $\epsilon$-secure. Also, Alice (Bob) could obtain $k_{\rm A}$ ($k_{\rm B}$) by using the reconstruct protocol of a VSS (see Appendix~\ref{tool}). That is, Alice (Bob) could use majority voting to obtain the shares $k_{{\rm A}ij}$ and $k_{{\rm B}ij'}$ of $k_{\rm A}$ ($k_{\rm B}$) from $k_{{\rm A}i'ij}$ ($k_{{\rm B}i''ij'}$) for all $i=1,\ldots,M$ and $j=1,\ldots,q$ ($j'=1,\ldots,q'$), and she (he) calculates $k_{\rm A}=\oplus_{i=1}^M\oplus_{j=1}^q k_{{\rm A}ij}$ ($k_{\rm B}=\oplus_{i=1}^M\oplus_{j'=1}^{q'} k_{{\rm B}ij'}$) where $q$ ($q'$) is the total number of shares of $k_{{\rm A}i}$ ($k_{{\rm B}i}$) for each $i$. 

Our results are summarised in the following Claim, whose proof is direct from the definition of {\it Protocol}~$3$:
\\

\noindent{\bf Claim 3.} {\it Suppose that Alice and Bob have $n$ pairs of QKD modules and Alice (Bob) has $s$ ($r$) classical post-processing units. Suppose that up to $t<n$ pairs of QKD modules, up to $t'<s/3$ classical post-processing units of Alice, and up to $t''<r/3$ classical post-processing units of Bob could be corrupted. Let $M\leq{}n$ denote the number of pairs of QKD modules that do not abort and whose raw key can be transformed into a supposedly $(\epsilon/n)$-secure key, and let $N$ bits be the length of such key. Then Protocol $3$ allows Alice and Bob to distill an $\epsilon$-secure key of length $(M-t)\times{}N$ bits. Moreover, the re-use of the devices does not compromise the security of the keys distilled in previous QKD runs.}
\\

We remark that if we disregard the cost of authenticating the classical channels between Alice and Bob's classical post-processing units, {\it Protocol}~$3$ is optimal with respect to the resulting secret key length. The argument follows directly from that used in Sec.~\ref{ups}, where we showed that if the classical post-processing units are trusted the secret key rate is upper bounded by $(M-t)\times{}N$ bits. So, in the presence of corrupted classical post-processing units this upper bound also trivially holds.

\section{Discussion and conclusions}

Security proofs of QKD assume that there are no covert channels and the classical post-processing units are trusted. Unfortunately, however, both assumptions are very hard, if not impossible, to guarantee in practice. 

Indeed, memory attacks~\cite{mem_attack} constitute a fundamental practical threat to the security of both DI-QKD and non-DI-QKD. They highlight that quantum mechanics alone is not enough to guarantee the security of practical QKD realisations but, for this, one needs to resort to additional assumptions. Also, recent results on Trojan Horse attacks~\cite{covert2,covert2b,covert2c,covert3,military1,military2} against conventional cryptographic systems underline the vulnerabilities of the classical post-processing units, and this threat is expected to only rise with time.

In this paper we have introduced a simple solution to overcome these two fundamental security problems and restore the security of QKD. The price to pay is that now Alice and Bob need to have various QKD modules and classical post-processing units at their disposal, bought for example from different vendors. Given that there is at least one pair of honest QKD modules and that the number of corrupted classical post-processing units is less than one third of them, we have shown how VSS schemes, together with privacy amplification techniques, could be used to re-establish the security of QKD. 

Indeed, VSS and secret sharing techniques have been used previously in quantum information~\cite{extra1,extra1b,extra2}. For instance, the authors of~\cite{extra1,extra1b} proposed a quantum version of VSS to achieve secure multiparty quantum computation, while in~\cite{extra2} classical secret sharing schemes are combined with QKD to achieve information-theoretically secure distributed storage systems.

A key insight of our paper is very simple yet potentially very useful: the typical classical post-processing in QKD only involves operations which are ``linear'' in nature, and thus they could be easily implemented in a distributed setting by acting on data shares from say a linear VSS scheme.

To illustrate our results, we have proposed specific protocols for three scenarios of practical interest. They assume that either the QKD modules, the classical-port-processing units, or both of them together could be corrupted. Remarkably, if we disregard the cost of classical authentication, all these protocols are optimal with respect to the secret key rate. They use the VSS scheme introduced in~\cite{maurer_smpc}, which is very simple to implement. Its main drawback is, however, that, for a given number of corrupted parties, the number of shares grows exponentially with the total number of parties. Nevertheless, for a small number of parties (which is the scenario that we are interested in QKD), the protocol is efficient in terms of computational complexity. Also, we remark that there exist efficient three-round VSS protocols where the computation and communication required is polynomial in the total number of parties~\cite{eff_VSS}. Moreover, these schemes use a minimum number of communication rounds~\cite{gen}, and they could also be used here. It would be interesting to further investigate the most resource efficient protocols to be used in the QKD framework.

\section{Acknowledgements}
We acknowledge support from the Galician Regional Government (consolidation of Research Units: AtlantTIC), the Spanish Ministry of Economy and Competitiveness (MINECO), the Fondo Europeo de Desarrollo Regional (FEDER) through Grant No.~TEC2014-54898-R, the European Commission (Project QCALL), NSERC, CFI and ORF.

\appendix

\section{Secure multiparty computation toolbox}\label{tool}
Here we briefly introduce some definitions and cryptographic protocols that are used in the main text; they are mainly taken from~\cite{book_smc,maurer_smpc}. 

We consider a scenario with a dealer and $n$ parties, and we suppose a threshold active adversary structure where Eve can actively corrupt the dealer and up to $t$ of the parties. Active corruption means that Eve can fully control the corrupted parties whose behaviour can deviate arbitrarily from the protocol's prescriptions. We refer the reader to Appendix~\ref{ap_struct} for the modeling of more general adversary structures. For simplicity, we assume that all messages are binary strings and the symbol $\oplus$ below denotes bit-wise XOR or bit-wise addition modulo $2$. We remark, however, that the the protocols below work as well over any finite field or ring.

In this scenario, a $n$-out-of-$q$ threshold secret sharing (SS) scheme~\cite{ss1,ss2} is a protocol that allows the dealer to split a message $m$ between $n$ parties such that, if he is honest, any group of $q$ or more parties can collaborate to reconstruct $m$ but no group with less than $q$ parties can obtain any information about $m$. If $n=q$, this could be achieved by splitting $m$ into a random sum of $q$ shares $m_i$. That is, one selects the first $q-1$ shares $m_i$ of $m$ at random, and then chooses $m_q=m\oplus{}m_1\oplus\ldots\oplus{}m_{q-1}$~\cite{book_smc}. 

A drawback of SS schemes is that they do not guarantee the consistency of the shares, which is essential to assure the correctness of the keys delivered by the QKD protocols in the main text. That is, during the reconstruct phase of a SS scheme, corrupted parties could send different $m_i$ to the honest parties such that they obtain different values for $m$. This problem can be solved with verifiable secret sharing (VSS) schemes~\cite{verifiable1,verifiable2}, which distribute $m_i$ in a redundant manner such that the honest parties can use error correction to obtain the correct values. Indeed, provided that the necessary and sufficient conditon $t<n/3$ is satisfied, a VSS scheme guarantees that there exists a well-defined $m$ that all honest parties obtain from their shares~\cite{maurer_smpc,smc1,smc2}. 

\begin{figure}
  \includegraphics[width=0.83\columnwidth]{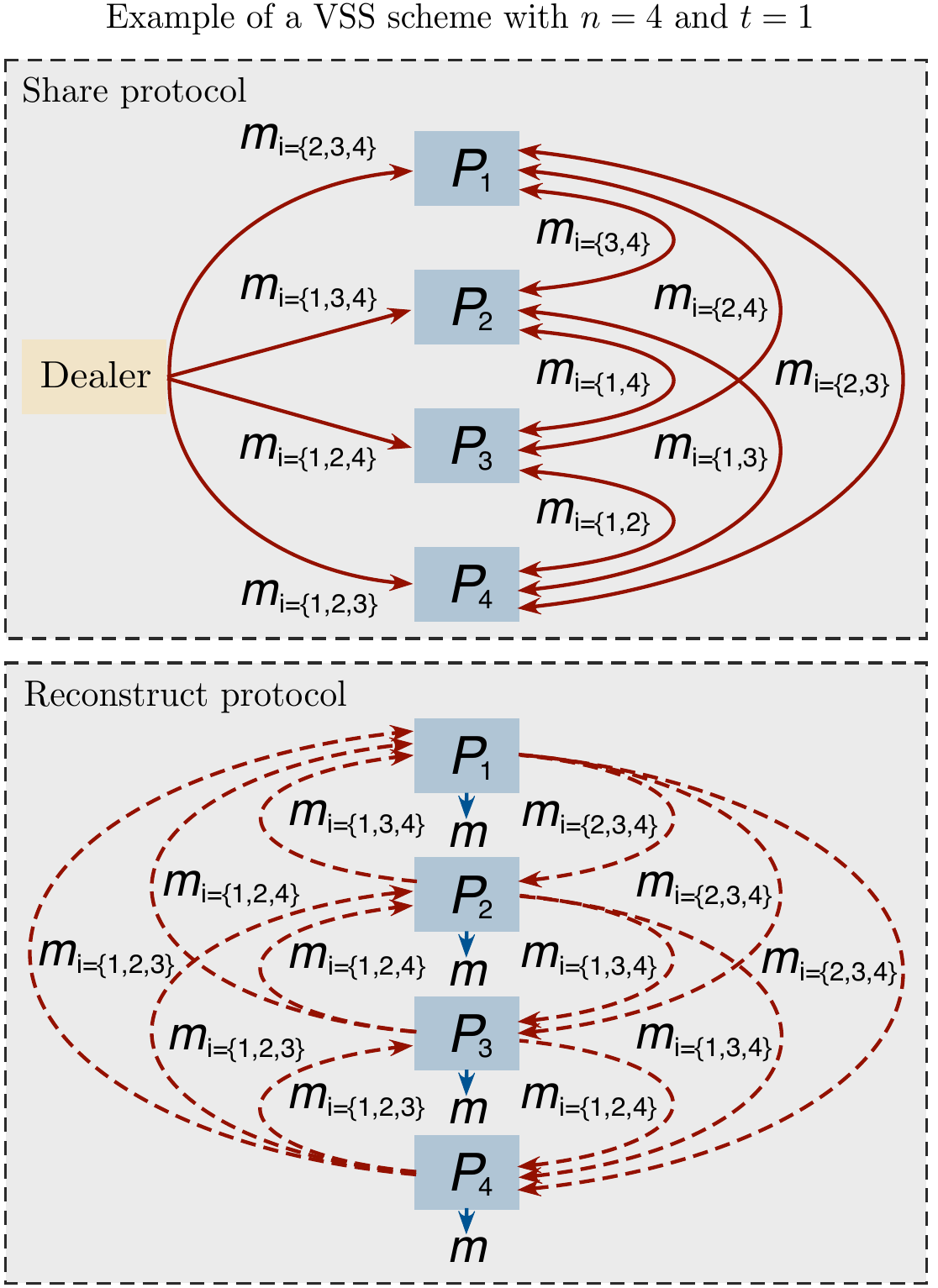}
\caption{Schematic representation of the share and the reconstruct protocols of the VSS scheme introduced in~\cite{maurer_smpc} for the case $n=4$ and $t=1$. The sets $\sigma_i$ are given by $\sigma_1=\{P_2,P_3,P_4\}$, $\sigma_2=\{P_1,P_3,P_4\}$, $\sigma_3=\{P_1,P_2,P_4\}$ and $\sigma_4=\{P_1,P_2,P_3\}$, with $P_i$ denoting the $i$-th party. In the share protocol, the dealer uses a $4$-out-of-$4$ SS scheme to split the message $m$ into shares $m_i$, with $i=1,\ldots,4$. Then, he sends each $m_i$ over a secure channel to all parties in $\sigma_i$. For instance, he sends $m_1$ to $P_2$, $P_3$, and $P_4$, and similarly for the other shares. Afterward, all parties in $\sigma_i$ send each other their shares $m_i$ to check that they are indeed equal. If no inconsistency is found then, in the reconstruct protocol, each party sends all his shares to all the other parties over an authenticated channel. Finally, each party uses majority voting to obtain $m_i$ $\forall i$ and then reconstructs $m=\oplus_{i=1}^{4} m_i$. In the figure, $m_{i=\{x,y,z\}}$ and $m_{i=\{x,y\}}$ denote, respectively, the bit strings $\{m_x,m_y,m_z\}$ and $\{m_x,m_y\}$ with $x,y,z\in\{1,2,3,4\}$, the red solid lines represent secure classical channels ({\it i.e.}, channels which guarantee both secrecy and authentication), the red dashed lines are authenticated classical channels, and the blue solid lines refer to the output of the reconstruct protocol. 
}
\label{fig4}
\end{figure}

The share and reconstruct protocols of a VSS scheme satisfy three conditions. First, independently of whether or not the dealer is honest, if the share protocol is successful then the reconstruct protocol delivers the same $m$ to all the honest parties. Second, if the dealer is honest, the value of the reconstructed $m$ coincides with that provided by the dealer. And third, if the dealer is honest, the information obtained by any set of $t$ of less parties after the share (reconstruct) protocol is independent of any previous information that they held before the protocol (is just the reconstructed bit string $m$). Below we present a simple VSS scheme that builds on the $q$-out-of-$q$ threshold SS protocol above~\cite{maurer_smpc}, and which we use in {\it Protocols}~$2$ and $3$ in the main text. Importantly, given that $t<n/3$, this scheme provides information-theoretic security~\cite{maurer_smpc}. See Fig.~\ref{fig4} for a graphical representation of its share and reconstruct protocols. 
\\

\noindent{}Share protocol: 
\begin{enumerate}
\item The dealer uses a $q$-out-of-$q$ SS scheme to split $m$ into $q={n \choose n-t}$ shares $m_i$, with $i=1,\ldots,q$.
\item Let $\{\sigma_1,\ldots,\sigma_q\}$ denote all $(n-t)$-combinations of the set of $n$ parties. Then, for each $i=1,\ldots,q$, the dealer sends $m_i$ over a secure channel ({\it i.e.}, a channel that provides secrecy and authentication) to each party in $\sigma_i$. If a party does not receive his share, he takes as default share say a zero bit string.
\item All pairs of parties in $\sigma_i$ send each other their shares $m_i$ over a secure channel to check if they are indeed equal. If an inconsistency is found, they complain using a broadcast channel.
\item If a complaint is raised in $\sigma_i$, the dealer broadcasts $m_i$ to all parties and they accept the share received. Otherwise, the protocol aborts.
\end{enumerate}

\noindent{}Reconstruct protocol: 
\begin{enumerate}
\item All pairs of parties send each other their shares over an authenticated channel.
\item Each party uses majority voting to reconstruct the shares $m_i$ $\forall i$, and then obtains $m=\oplus_{i=1}^{q} m_i$.
\end{enumerate}

From the description above, it is guaranteed that when the share protocol is successful ({\it i.e.}, it does not abort), all the honest parties who received the $i$-th share of $m$ obtain exactly the same bit string $m_i$. Also, this protocol assures that any share $m_i$ of $m$ is distributed to at least $2t+1$ different parties. This is so because this is the minimum size of any set $\sigma_i$. This means, in particular, that, since the number of corrupted parties is at most $t$, the use of a decision rule based on majority voting in the reconstruct protocol permits all the honest parties to obtain the same fixed $m_i$ for all $i$. Moreover, it is straightforward to show that when the dealer is honest, the reconstructed message $m$ is equal to his original message. Furthermore, we have that $m$ is only revealed to the parties once the reconstruct phase ends. This is so because at least one bit string $m_i$ is only shared by honest parties since there is at least one set $\sigma_i$ which does not contain any corrupted party. Also, note that if a complaint is raised in a certain $\sigma_i$ during the share protocol, the fact that the dealer broadcast $m_i$ to all parties does not violate secrecy. This is so because a complaint can only occur if either the dealer is corrupted or $\sigma_i$ contains at least one corrupted player, hence the adversary knew $m_i$ already.

We remark that the broadcast channel which is required in steps $3$ and $4$ of the share protocol can be a simulated channel. Indeed, given that $t<n/3$, there exist efficient poly($n$) protocols that can simulate a broadcast channel with information-theoretic security in an optimal number of $t+1$ communication rounds~\cite{garay,fischer}. Furthermore, if a physical broadcast channel is actually available, there exist efficient information-theoretically secure VSS schemes that only require a majority of honest parties ({\it i.e.}, $t<n/2$) and which could also be used in this context~\cite{tal}.

Next we present a simple scheme to generate a common perfectly unbiased random $l$-bit string (RBS) $r$ between $n$ parties when up to $t<n/3$ of them could be corrupted. It follows directly from VSS~\cite{maurer_smpc,smc1,smc2}. For convenience, we call it the RBS protocol. We use it to randomly select universal$_2$ hash functions in {\it Protocols}~$2$ and $3$ in the main text, where we cannot assume the existence of an external honest dealer which provides them to the QKD devices.~The RBS scheme allows mutually untrusted parties to generate and share random numbers through discussions. 
\\

\noindent{}RBS protocol: 
\begin{enumerate}
\item Say each of the first $t+1$ parties produces locally a random $l$-bit string $r_i$ and sends it to all the other parties using the share protocol above.
\item Each party uses a broadcast channel to confirm that they have received all their shares from the first $t+1$ parties. Otherwise, the protocol aborts.
\item All parties use the reconstruct protocol above to obtain $r_i$ for all $i=1,\ldots,t+1$. Afterward, each of them calculates locally $r=\oplus_{i=1}^{t+1}r_{i}$.
\end{enumerate}

It is straightforward to show that this protocol guarantees that all honest parties share a perfectly unbiased random bit string $r$. The use of the share and the reconstruct protocols of a VSS scheme assures that all honest parties reconstruct the same $r_i$ $\forall{}i$ and thus the same $r$. In addition, step $2$ of the protocol guarantees that the first $t+1$ parties generate and distribute their strings $r_i$ before knowing the strings of the other parties. Moreover, since the number of corrupted parties is at most $t$, we have that at least one honest party generates a truly random bit string $r_i$, and thus $r$ is also random. 

\section{Protocol~$2$}\label{app_P2}

Here we present the different steps of {\it Protocol}~$2$ in detail. For concreteness, whenever we refer to the share and reconstruct protocols of a VSS scheme we mean those presented in Appendix~\ref{tool}, which have been introduced in~\cite{maurer_smpc}.

Also, to simplify the discussion, in {\it Protocol}~$2$ we consider the case where $p_{\rm A,info}$ and $p_{\rm B,info}$ determine the sifting procedure of the QKD scheme in a deterministic way. That is, there is no {\it random} post-selection of data from the raw key. In addition, we assume that Alice and Bob do not estimate the actual QBER but they apply error correction for a pre-fixed QBER value followed by an error verification step. However, we remark that {\it Protocol}~$2$ could be adapted to cover also these two scenarios. 
\\

\begin{enumerate}
\item {\it Generation and distribution of shares of raw keys and protocol information}: QKD$_{\rm A}$ and QKD$_{\rm B}$ obtain, respectively, the raw keys $k'_{\rm A}$ and $k'_{\rm B}$ and the protocol information $p_{\rm A,info}$ and $p_{\rm B,info}$, or the abort symbol $\perp$. If the result is different from $\perp$, QKD$_{\rm A}$ uses the share protocol of a VSS scheme to create $q={s \choose s-t'}$ shares of $k'_{\rm A}$ and distributes them among the CP$_{{\rm A}i}$, with $i=1,\ldots,s$. Likewise, QKD$_{\rm B}$ creates $q'={r \choose r-t''}$ shares of $k'_{\rm B}$ and distributes them among the CP$_{{\rm B}i'}$, with $i'=1,\ldots,r$. Let $k'_{{\rm A}ij}$ ($k'_{{\rm B}i'j'}$) be the $j$-th ($j'$-th) share of $k'_{\rm A}$ ($k'_{\rm B}$) received by CP$_{{\rm A}i}$ (CP$_{{\rm B}i'}$), with $j=1,\ldots,q$ ($j'=1,\ldots,q'$). Also, QKD$_{\rm A}$ (QKD$_{\rm B}$) sends $p_{\rm A,info}$ ($p_{\rm B,info}$) to all CP$_{{\rm A}i}$ (CP$_{{\rm B}i'}$). Since by assumption QKD$_{\rm A}$ (QKD$_{\rm B}$) is honest, all CP$_{{\rm A}i}$ (CP$_{{\rm B}i'}$) receive the same $p_{\rm A,info}$ ($p_{\rm B,info}$) and the shares $k'_{{\rm A}ij}$ ($k'_{{\rm B}i'j'}$) are equal for all $i$ ($i'$). Next, say the first $2t''+1$ CP$_{{\rm B}i'}$ send $p_{\rm B,info}$ to all CP$_{{\rm A}i}$. Likewise, say the first $2t'+1$ CP$_{{\rm A}i}$ send $p_{\rm A,info}$ (for the detected events) to all CP$_{{\rm B}i'}$. Each CP$_{{\rm A}i}$ (CP$_{{\rm B}i'}$) uses majority voting to determine $p_{\rm B,info}$ ($p_{\rm A,info}$) from the information received. Note that since by assumption the number of corrupted units CP$_{{\rm A}i}$ (CP$_{{\rm B}i'}$) is at most $t'$ ($t''$), $2t'+1$ ($2t''+1$) copies of $p_{\rm A,info}$ ($p_{\rm B,info}$) is enough for the honest parties to be able to reconstruct the correct value of these bit strings by using majority voting.
\item {\it Sifting}: Each CP$_{{\rm A}i}$ uses $p_{\rm A,info}$ and $p_{\rm B,info}$ to obtain two bit strings, $k_{\rm A{\it ij}, key}$ and $k_{\rm A{\it ij}, est}$, from $k'_{{\rm A}ij}$. The former (latter) bit string is the part of $k'_{{\rm A}ij}$ that is used for key generation (parameter estimation). Likewise, Bob's CP$_{{\rm B}i'}$ do the same with $k'_{{\rm B}i'j'}$ and obtain $k_{\rm B{\it i'j'}, key}$ and $k_{\rm B{\it i'j'}, est}$.
\item {\it Parameter estimation}: All CP$_{{\rm A}i}$ and CP$_{{\rm B}i'}$ use the reconstruct protocol of a VSS scheme to obtain both $k_{\rm A, est}$ and $k_{\rm B, est}$, which are the parts of $k'_{\rm A}$ and $k'_{\rm B}$ that are used for parameter estimation. For this, they send each other their shares $k_{\rm A{\it ij}, est}$ and $k_{\rm B{\it i'j'}, est}$, and each of them uses majority voting to obtain $k_{\rm A{\it j}, est}$ and $k_{\rm B{\it j'}, est}$ for all $j=1,\ldots,q$ and $j'=1,\ldots,q'$. Afterward, they calculate $k_{\rm A, est}=\oplus_{j=1}^qk_{\rm A{\it j}, est}$ and $k_{\rm B, est}=\oplus_{j'=1}^{q'}k_{\rm B{\it j'}, est}$. With $p_{\rm A,info}$, $p_{\rm B,info}$, $k_{\rm A, est}$ and $k_{\rm B, est}$, each CP$_{{\rm A}i}$ and CP$_{{\rm B}i'}$ performs locally the parameter estimation step of the protocol ({\it e.g.}, they estimate the phase error rate). If the estimated values exceed certain tolerated values, they abort.
\item {\it Error correction}: The CP$_{{\rm A}i}$ and CP$_{{\rm B}i'}$ perform error correction (for a pre-fixed QBER value) on the parts of $k'_{\rm A}$ and $k'_{\rm B}$ that are used for key distillation, which we denote by $k_{\rm A, key}$ and $k_{\rm B, key}$, by acting on their shares $k_{\rm A{\it ij}, key}$ and $k_{\rm B{\it i'j'}, key}$ respectively. For this, each CP$_{{\rm A}i}$ (CP$_{{\rm B}i'}$) applies certain matrices $M_{\rm EC}$ to $k_{\rm A{\it ij}, key}$ ($k_{\rm B{\it i'j'}, key}$) to obtain $s_{{\rm A}ij}=M_{\rm EC}k_{\rm A{\it ij}, key}$ ($s_{{\rm B}i'j'}=M_{\rm EC}k_{\rm B{\it i'j'}, key}$). Afterward, they use the reconstruct protocol of a VSS scheme to guarantee that all CP$_{{\rm B}i'}$ obtain $s_{\rm A}=M_{\rm EC}k_{\rm A, key}$ and $s_{\rm B}=M_{\rm EC}k_{\rm B, key}$. That is, all CP$_{{\rm A}i}$ and CP$_{{\rm B}i'}$ first send to all the classical post-processing units at Bob's side the bit strings $s_{{\rm A}ij}$ and $s_{{\rm B}i'j'}$. Then, each of Bob's CP units uses majority voting to reconstruct locally $s_{{\rm A}j}$ and $s_{{\rm B}j'}$, for all $j$ and $j'$, from $s_{{\rm A}ij}$ and $s_{{\rm B}i'j'}$. Finally, they obtain $s_{\rm A}=\oplus_{j=1}^qs_{{\rm A}j}$ and $s_{\rm B}=\oplus_{j'=1}^{q'}s_{{\rm B}j'}$. Next, Bob corrects $k_{\rm B, key}$. For this, say all CP$_{{\rm B}i'}$ which have the $j'$-th share $k_{\rm B{\it i'j'}, key}$ for a pre-fixed index $j'=1,\ldots,q'$, flip certain bits of this share depending on the actual values of $s_{\rm A}$ and $s_{\rm B}$. This whole process is repeated until the error correction procedure ends. Let ${\hat k}_{\rm A{\it ij}, key}$ and ${\hat k}_{\rm B{\it i'j'}, key}$ denote the shares $k_{\rm A{\it ij}, key}$ and $k_{\rm B{\it i'j'}, key}$ after error correction, and let leak$_{\rm EC}$ bits be the syndrome information interchanged between Alice and Bob during this step. That is, ${\hat k}_{\rm A{\it ij}, key}$ and ${\hat k}_{\rm B{\it i'j'}, key}$ are actually equal to $k_{\rm A{\it ij}, key}$ and $k_{\rm B{\it i'j'}, key}$ except for the bit strings $k_{\rm B{\it i'j'}, key}$ whose bits have been flipped during error correction.
\item {\it Error verification}: All CP$_{{\rm A}i}$ and CP$_{{\rm B}i'}$ check that the error correction step was indeed successful. For this, the CP$_{{\rm A}i}$ use the RBS scheme introduced in Appendix~\ref{tool} to randomly select a universal$_2$ hash function, $h_{\rm V}$. Then, they compute a hash $h_{{\rm A}ij}=h_{\rm V}({\hat k}_{\rm A{\it ij}, key})$ of length $\lceil\log_2{(4/\epsilon_{\rm cor})}\rceil$ bits, and say the first $2t'+1$ CP$_{{\rm A}i}$ send the hash function to all CP$_{{\rm B}i'}$. Bob's CP units reconstruct the hash function by using majority voting and then they calculate $h_{{\rm B}i'j'}=h_{\rm V}({\hat k}_{\rm B{\it i'j'}, key})$. Afterward, all CP$_{{\rm A}i}$ and CP$_{{\rm B}i'}$ use the reconstruct protocol of a VSS scheme to obtain $h_{{\rm A}}=\oplus_{j=1}^qh_{{\rm A}j}$ and $h_{{\rm B}}=\oplus_{j'=1}^{q'}h_{{\rm B}j'}$ from $h_{{\rm A}ij}$ and $h_{{\rm B}i'j'}$. That is, they send each other $h_{{\rm A}ij}$ and $h_{{\rm B}i'j'}$ and they use majority voting to determine $h_{{\rm A}j}$ and $h_{{\rm B}j'}$, for all $j$ and $j'$, from $h_{{\rm A}ij}$ and $h_{{\rm B}i'j'}$. Finally, each of them checks locally whether or not $h_{\rm A}=h_{\rm B}$. If they are not equal, they abort. In so doing, we have that the bit strings ${\hat k}_{\rm A, key}=\oplus_{j=1}^q {\hat k}_{\rm A{\it j}, key}$ and ${\hat k}_{\rm B, key}=\oplus_{j'=1}^{q'} {\hat k}_{\rm B{\it j'}, key}$ are equal except for a minuscule probability $\epsilon_{\rm cor}$, where ${\hat k}_{\rm A{\it j}, key}$ (${\hat k}_{\rm B{\it j'}, key}$) are obtained from ${\hat k}_{\rm A{\it ij}, key}$ (${\hat k}_{\rm B{\it i'j'}, key}$) by using majority voting.
\item {\it Generation of shares of an $\epsilon$-secure key}: All CP$_{{\rm A}i}$ and CP$_{{\rm B}i'}$ extract from ${\hat k}_{\rm A, key}$ and ${\hat k}_{\rm B, key}$ the shares of an $\epsilon_{\rm sec}$-secret key, $k_{\rm A}$ and $k_{\rm B}$. For this, the CP$_{{\rm A}i}$ use the RBS scheme to randomly select a proper universal$_2$ hash function, $h_{\rm P}$. Next, they obtain $k_{{\rm A}ij}=h_{\rm P}({\hat k}_{\rm A{\it ij}, key})$ and say the first $2t'+1$ CP$_{{\rm A}i}$ send $h_{\rm P}$ to all CP$_{{\rm B}i'}$. Bob's CP units use majority voting to determine $h_{\rm P}$ from the information received and they calculate $k_{{\rm B}i'j'}=h_{\rm P}({\hat k}_{\rm B{\it i'j'}, key})$. The function $h_{\rm P}$ removes Eve's information from ${\hat k}_{\rm A, key}$, which includes the syndrome information leak$_{\rm EC}$ disclosed during error correction, the hash value of length $\lceil\log_2{(4/\epsilon_{\rm cor})}\rceil$ bits disclosed during error verification, and Eve's information about the key according to the estimated phase error rate.
\end{enumerate}

As stated in Sec.~\ref{ups2}, when $t'<M_{\rm A}/3$ and $t''<M_{\rm B}/3$, where $M_{\rm A}$ ($M_{\rm B}$) is the number of CP$_{{\rm A}i}$ (CP$_{{\rm B}i'}$) that do not abort, $k_{{\rm A}ij}$ and $k_{{\rm B}i'j'}$ are shares of an $\epsilon$-secure key, $k_{\rm A}$ and $k_{\rm B}$. This is so because the condition $t'<M_{\rm A}/3$ (or, equivalently, $s-t'-(s-M_{\rm A})>2t'$) guarantees that for all $j=1,\ldots,q$, there are at least $2t'+1$ units CP$_{{\rm A}i}$ which send shares $k_{{\rm A}ij}$ to Alice. To see this, note that each share $k_{{\rm A}j}$, for all $j$, is held by $s-t'$ units CP$_{{\rm A}i}$, and by assumption we have that at most $s-M_{\rm A}$ of them could have aborted. A similar argument applies to the condition $t''<M_{\rm B}/3$.

To reconstruct $k_{\rm A}$ and $k_{\rm B}$, Alice and Bob can use majority voting to obtain $k_{{\rm A}j}$ and $k_{{\rm B}j'}$ from $k_{{\rm A}ij}$ and $k_{{\rm B}i'j'}$, respectively, and afterward they calculate $k_{\rm A}=\oplus_{j=1}^q k_{{\rm A}j}$ and $k_{\rm B}=\oplus_{j'=1}^{q'} k_{{\rm B}j'}$.

\section{Alternative solution for QKD with honest QKD modules and malicious classical post-processing units}\label{alter}

In this Appendix we present a conceptually simple, although less efficient, solution than {\it Protocol}~$2$ for the case where $r=s$. 

The main idea runs as follows. First, QKD$_{\rm A}$ and QKD$_{\rm B}$ perform $s$ independent QKD sessions, each of them is realised with a different pair of units CP$_{{\rm A}i}$ and CP$_{{\rm B}i}$ to generate a supposedly $(\epsilon/s)$-secure key, $k_{{\rm A}i}$ and $k_{{\rm B}i}$, or the abort symbol $\perp_i$. For easy of illustration, we shall assume that the length of each $k_{{\rm A}i}$ and $k_{{\rm B}i}$ is $N$ bits for all $i$. Of course, if say CP$_{{\rm A}i}$ and/or CP$_{{\rm B}i}$ is corrupted then we have that $k_{{\rm A}i}$ and $k_{{\rm B}i}$ could be compromised and known to Eve. Then, in a second step, the keys $k_{{\rm A}i}$ and $k_{{\rm B}i}$ are concatenated to form ${\hat k}_{\rm A}=[k_{{\rm A}1},\ldots,k_{{\rm A}M}]$ and ${\hat k}_{\rm B}=[k_{{\rm B}1},\ldots,k_{{\rm B}M}]$, where $M$ denotes the number of keys $k_{{\rm A}i}$ and $k_{{\rm B}i}$ which are different from the abort symbol. Finally, we apply error verification and privacy amplification to ${\hat k}_{\rm A}$ and ${\hat k}_{\rm B}$ to obtain an $\epsilon$-secure key, $k_{\rm A}$ and $k_{\rm B}$. Importantly, this last step is performed by the classical post-processing units in a distributed setting by acting only on shares of ${\hat k}_{\rm A}$ and ${\hat k}_{\rm B}$. Below we describe the different steps of the protocol in more detail.
\\

\noindent{\it Alternative solution to Protocol $2$}: 
\begin{enumerate}
\item {\it Generation of $(\epsilon/s)$-secure keys}: QKD$_{\rm A}$ and QKD$_{\rm B}$ perform $s$ independent QKD sessions, each of which with a different pair of units CP$_{{\rm A}i}$ and CP$_{{\rm B}i}$, with $i=1,\ldots,s$, to obtain the bit strings $k_{{\rm A}i}$ and $k_{{\rm B}i}$, which are supposed to be $(\epsilon/s)$-secure, or the abort symbol $\perp_i$. 
\item {\it Distribution of shares of $(\epsilon/s)$-secure keys}: Each CP$_{{\rm A}i}$ sends $k_{{\rm A}i}$ to the other classical post-processing units at Alice's side by using the share protocol of a VSS scheme, and all CP$_{{\rm A}i'}$ confirm to each other that they have received their shares. Let $k'_{{\rm A}i'ij}$ be the $j$-th share of $k_{{\rm A}i}$ received by CP$_{{\rm A}i'}$. Likewise, the units CP$_{{\rm B}i}$ act similarly with $k_{{\rm B}i}$. Let $k'_{{\rm B}i'ij}$ be the $j$-th share of $k_{{\rm B}i}$ received by CP$_{{\rm B}i'}$. Each CP$_{{\rm A}i'}$ defines locally the bit strings $k''_{{\rm A}i'ij}=[\vec{0}_1,\ldots, \vec{0}_{i-1}, k'_{{\rm A}i'ij},\vec{0}_{i+1},\ldots, \vec{0}_{M}]$, where $\vec{0}$ is the $N$-bit zero vector and $M$ is the number of keys $k_{{\rm A}i}$ and $k_{{\rm B}i}$ which are different from $\perp_i$. Likewise, the CP$_{{\rm B}i'}$ form $k''_{{\rm B}i'ij}$ from $k'_{{\rm B}i'ij}$. 
\item {\it Error verification}: The ${\rm CP}_{{\rm A}i'}$ use the RBS scheme to randomly select a universal$_2$ hash function, $h_{\rm V}$. Then, each of them computes locally a hash $h_{{\rm A}i'ij}=h_{\rm V}(k''_{{\rm A}i'ij})$ of length $\lceil\log_2{(4/\epsilon_{\rm cor})}\rceil$ bits for all its bit strings $k''_{{\rm A}i'ij}$, and say the first $2t'+1$  ${\rm CP}_{{\rm A}i'}$ send the hash function to all CP units at Bob's side. Each ${\rm CP}_{{\rm B}i'}$ reconstructs locally the hash function by using majority voting and obtains $h_{{\rm B}i'ij}=h_{\rm V}(k''_{{\rm B}i'ij})$ for all its bit strings $k''_{{\rm B}i'ij}$. Next, all ${\rm CP}_{{\rm A}i'}$ and ${\rm CP}_{{\rm B}i'}$ use the reconstruct protocol of a VSS scheme to obtain both $h_{{\rm A}}=\oplus_{i=1}^M\oplus_{j=1}^qh_{{\rm A}ij}$ and $h_{{\rm B}}=\oplus_{i=1}^M\oplus_{j=1}^qh_{{\rm B}ij}$. That is, they send each other the bit strings $h_{{\rm A}i'ij}$ and $h_{{\rm B}i'ij}$, and each of them uses majority voting to obtain $h_{{\rm A}ij}$ and $h_{{\rm B}ij}$ from $h_{{\rm A}i'ij}$ and $h_{{\rm B}i'ij}$. Finally, each ${\rm CP}_{{\rm A}i'}$ and ${\rm CP}_{{\rm B}i'}$ checks locally if $h_{\rm A}=h_{\rm B}$. If they are not equal, they output the abort symbol $\perp_{i'}$. If they are equal, they proceed to the next step. This error verification step guarantees that $k''_{\rm A}=\oplus_{i=1}^M\oplus_{j=1}^qk''_{{\rm A}ij}$ and $k''_{\rm B}=\oplus_{i=1}^M\oplus_{j=1}^qk''_{{\rm B}ij}$ are equal except for a minuscule probability $\epsilon_{\rm cor}$, where $k''_{{\rm A}ij}$ ($k''_{{\rm B}ij}$) denote the bit strings that could be obtained from $k''_{{\rm A}i'ij}$ ($k''_{{\rm B}i'ij}$) by using majority voting.
\item {\it Generation of shares of an $\epsilon$-secure key}: The ${\rm CP}_{{\rm A}i'}$ use the RBS scheme to randomly select a universal$_2$ hash function, $h_{\rm P}$, and they compute $k_{{\rm A}i'ij}=h_{\rm P}(k''_{{\rm A}i'ij})$. Then, say the first $2t'+1$  ${\rm CP}_{{\rm A}i'}$ send $h_{\rm P}$ to all CP at Bob's side which reconstruct locally the hash function by using majority voting, and each ${\rm CP}_{{\rm B}i'}$ computes $k_{{\rm B}i'ij}=h_{\rm P}(k''_{{\rm B}i'ij})$. The function $h_{\rm P}$ maps each $(M\times{}N)$-bit string $k''_{{\rm A}i'ij}$ ($k''_{{\rm B}i'ij}$) to a shorter bit string $k_{{\rm A}i'ij}$ ($k_{{\rm B}i'ij}$) of size $(M-2t')\times{}N-\lceil\log_2{(4/\epsilon_{\rm cor})}\rceil$ bits.
\end{enumerate}

The reason for reducing the size of $k''_{{\rm A}i'ij}$ and $k''_{{\rm B}i'ij}$ by $2t'\times{}N$ bits in the last step of the protocol is due to the following. In the worst-case scenario, we have that all corrupted CP$_{{\rm A}i}$ could be partnered with honest CP$_{{\rm B}i}$ (and vice versa). This means, in particular, that there could $2t'$ keys $k_{{\rm A}i}$ and $k_{{\rm B}i}$ which could be compromised and, more importantly, Alice and Bob cannot discard that these keys contribute to $k''_{{\rm A}i'ij}$ and $k''_{{\rm B}i'ij}$.

Given that $t'<M_{{\rm A}i}/3$ and $t''<M_{{\rm B}i}/3$ for all $i=1,\ldots,M$, where $M_{{\rm A}i}$ ($M_{{\rm B}i}$) denotes the number of CP$_{{\rm A}i'}$ (CP$_{{\rm B}i'}$) that do not produce $\perp_i$ but output post-processed shares, $k_{{\rm A}i'ij}$ ($k_{{\rm B}i'ij}$), from $k_{{\rm A}i}$ ($k_{{\rm B}i}$), then the final key, $k_{\rm A}$ and $k_{\rm B}$, is $\epsilon$-secure. Once again, to reconstruct it, Alice (Bob) can use majority voting to obtain $k_{{\rm A}ij}$ ($k_{{\rm B}ij}$) from $k_{{\rm A}i'ij}$ ($k_{{\rm B}i'ij}$) and then calculate $k_{\rm A}=\oplus_{i=1}^M\oplus_{j=1}^q k_{{\rm A}ij}$ ($k_{\rm B}=\oplus_{i=1}^M\oplus_{j=1}^q k_{{\rm B}ij}$). 

To conclude this Appendix, let us briefly compare the solution above with that provided by {\it Protocol}~$2$. For this, we first note that the approach above runs $s$ independent QKD sessions while {\it Protocol}~$2$ can distill an $\epsilon$-secure key from one single QKD run. The second main difference is related to the resulting secret key rate. To simplify our discussion, we shall assume, like above, that the length of each $k_{{\rm A}i}$ and $k_{{\rm B}i}$ is $N$ bits for all $i$. After running $s$ QKD sessions, the alternative protocol above can deliver a final key of length roughly $\approx(M-2t')\times{}N$ bits, while if we run {\it Protocol}~$2$ $s$ times the length of the final key would be roughly $s\times{}N$ secret bits. That is, even if we consider the best-case scenario for the alternative approach above ({\it i.e.}, $M=s$), {\it Protocol}~$2$ provides a secret key rate that is $\approx{}s/(s-2t')$ times higher than that provided by this alternative method.

\section{General adversary structures}\label{ap_struct}

In the main text we have considered the security of QKD against a so-called threshold active adversary structure. As we have already seen, active corruption means that there could exist a central adversary, Eve, who fully controls the behaviour of all the corrupted parties, which do not have to necessarily follow the prescriptions of the protocol. On the other hand, by a threshold adversary structure we refer to an adversary who can corrupt up to $t$ (but not more) of the parties. 

This is, however, a particular case of what is called a general mixed adversary structure~\cite{fitzi}. By mixed corruption we mean that some of the corrupted parties could also be passively (in contrast to actively) corrupted. Passive corruption indicates that the parties could leak all their information to the adversary, but otherwise they follow all the indications of the protocol correctly. General adversary structures, on the other hand, refer to the fact that the subsets that contain all the potentially corrupted parties could have an arbitrary distribution, {\it i.e.}, they do not need to consists on all possible combinations of up to $t$ parties. 

To model the corruption capability of a general adversary one can use a so-called $(\Sigma,\Omega)$-adversary structure. This is basically a set that contains all the potentially corruptible subsets of parties. More precisely, let $P$ denote the set of all parties, and let $\Sigma$ and $\Omega$ be structures for $P$ satisfying $\Sigma\subseteq\mathscr{P}(P)$ and $\Omega\subseteq\Sigma$ with $\mathscr{P}(P)$ being the power set of $P$. Here, a structure for $P$ means a subset $\Gamma$ of $\mathscr{P}(P)$ that is closed under taking subsets. That is, if $S\in\Gamma$ and $S'\subseteq{}S$ then $S'\in\Gamma$.  Then, a $(\Sigma,\Omega)$-adversary is an adversary that can passively (actively) corrupt a set $\sigma$ ($\omega$) of parties with $\sigma\cup{}\omega\in\Sigma$ and $\omega\in\Omega$. Below, for ease of notation, whenever we describe a structure, we only list its maximal sets. That is, it is implicitly understood (even if it is not explicitly written) that  its subsets also belong to the structure.

Next, we introduce the share and the reconstruct protocols of a VSS scheme~\cite{maurer_smpc} that provides information-theoretic security against a  general $(\Sigma,\Omega)$-adversary given that $P\notin\Sigma\sqcup\Omega\sqcup\Omega$, which can be proven to be a necessary and sufficient condition to achieve security in this framework~\cite{maurer_smpc}.  That is, the VSS below is optimal in this sense. Here, $\sqcup$ is an operation on structures defined as $\Gamma_1\sqcup\Gamma_2=\{S_1\cup{}S_2:S_1\in\Gamma_1, S_2\in\Gamma_2\}$. That is, $\Gamma_1\sqcup\Gamma_2$ is a structure that contains all unions of one element of $\Gamma_1$ and one element of $\Gamma_2$. If $\Sigma=\Omega$ ({\it i.e.}, when all corrupted parties are active), note that the condition $P\notin\Sigma\sqcup\Omega\sqcup\Omega$ coincides with that introduced in~\cite{smc1,smc2}.

Without loss of generality, below we assume that $\Sigma$ contains $q$ maximal sets $\sigma_i$, {\it i.e.}, $\Sigma=\{\sigma_1, \sigma_2, \ldots, \sigma_q\}$.
\\

\noindent{}Share protocol: 
\begin{enumerate}
\item The dealer uses a $q$-out-of-$q$ SS scheme to split the message $m$ into $q$ shares $m_i$, with $i=1,\ldots,q$.
\item For each $i=1,\ldots,q$, the dealer sends $m_i$ over a secure channel to each party in the set $\sigma'_i$, where $\sigma'_i$ is defined as the complement of $\sigma_i$. If a party does not receive his share, he takes as default share say a zero bit string.
\item All pairs of parties in $\sigma'_i$ send each other their shares $m_i$ over a secure channel to check that their shares are indeed equal. If an inconsistency is found, they complain using a broadcast channel.
\item If there is a complaint in $\sigma'_i$, the dealer broadcasts $m_i$ to all parties and they accept the share received. Otherwise, the protocol aborts.
\end{enumerate}

\noindent{}Reconstruct protocol: 
\begin{enumerate}
\item All parties send their shares to all other parties over an authenticated channel.
\item Each party reconstructs locally the shares $m_i$ $\forall i$, and obtains $m=\oplus_{i=1}^{q} m_i$. For this, let $m_{li}$ be the value for $m_i$ sent by the $l$-th party in $\sigma'_i$. Then, each party chooses the unique value $m_i$ such that there exists a $\omega\in\Omega$ satisfying $m_{li}=m_i$ for all $l\in\sigma'_i-\omega$.
\end{enumerate}

As in the case of the share protocol presented in the main text, the share protocol above also requires the availability of a broadcast channel. Fortunately, however, in this framework it is also possible to simulate such a channel by efficient protocols, with polynomial message and computation complexity, between the different parties~\cite{broad4}. For this, the requirement is that $P\notin\Omega\sqcup\Omega\sqcup\Omega$~\cite{maurer_smpc}. 

To conclude this Appendix, let us mention that the simple RBS protocol described in the main text can be straightforwardly adapted to be secure also against a general $(\Sigma,\Omega)$-adversary given that $P\notin\Sigma\sqcup\Omega\sqcup\Omega$. For this, we only need to make two modifications. First, we replace the share and reconstruct protocols with the ones described above. And, second, now we do not employ the first $t+1$ parties $P_i\in{}P$ to produce a random bit string each (see step $1$ of the RBS protocol in the main text). Instead, these random bit strings are produced by all the parties in one set $\rho\in\mathscr{P}(P)$ such that $\rho\nsubseteq\Sigma$. In so doing, we guarantee that there is at least one honest party that generates a random bit string. Here, one could take, for instance, the set $\rho$ with the minimum number of parties.

\section{QKD secure against general adversary structures}\label{gen_sec}

Finally, here we revisit briefly the three practical scenarios that we have considered in the main text, and we discuss how one could easily adapt {\it Protocols}~$1$, $2$ and $3$ to make them secure against a $(\Sigma,\Omega)$-adversary. For this, we shall assume that the structures $\Sigma$ and $\Omega$ are known to both Alice and Bob, which is a standard assumption in this framework. 

The case of {\it Protocol}~$1$ is rather simple, as it does not require any change. Since in that scenario one assumes that the classical post-processing units CP$_{\rm A}$ and CP$_{\rm B}$ are both honest, it is guaranteed that the concatenated bit strings $k'_{\rm A}$ and $k'_{B}$ are $\epsilon_{\rm cor}$-correct. Also, we have that Eve could know at most $t\times{}N$ bits of say $k'_{\rm A}$ except with probability $\epsilon_{\rm sec}$, where the parameter $t$ now refers to the size of the biggest set in $\Sigma$. That is, here $t$ denotes the maximum number of pairs of QKD modules that can be passively corrupted. Then, by applying a privacy amplification step to $k'_{\rm A}$ and $k'_{B}$, the units CP$_{\rm A}$ and CP$_{\rm B}$ can directly extract an $\epsilon_{\rm sec}$-secret key, $k_{\rm A}$ and $k_{B}$, of length $(M-t)\times{}N$ bits. In fact, one could even use a tighter bound here by selecting the parameter $t$ as the maximum size of all maximal sets in a structure $\Sigma'$ that is obtained from $\Sigma$ by removing from all of its subsets those pairs of QKD modules which have aborted.

The case of {\it Protocol}~$2$, on the other hand, requires a few minor modifications. First, now Alice and Bob have to use the share, reconstruct, and RBS protocols introduced above instead of the protocols presented in the main text. This implies that the $(\Sigma,\Omega)$-adversary structure has to satisfy both in Alice and Bob's side $P\notin\Sigma\sqcup\Omega\sqcup\Omega$ to guarantee security. Second, in steps $1$, $5$ and $6$ of {\it Protocol}~$2$, now we do not employ the first $2t'+1$ ($2t''+1$) units CP$_{{\rm A}i}$ (CP$_{{\rm B}i'}$) to reveal the information $p_{\rm A,info}$ ($p_{\rm B,info}$) as well as the hash functions $h_{\rm V}$ and $h_{\rm P}$. Instead, this information is revealed by all units in one set $\sigma'_i$, where $\sigma'_i$ is again the complement of $\sigma_i\in\Sigma$, both for Alice and Bob. One could take, for instance, the set $\sigma'_i$ with the minimum number of parties. Afterward, the information is reconstructed by using the second step of the reconstruct protocol introduced above. Note that such VSS scheme guarantees that, if all the parties in $\sigma'_i$ receive the same information, this information can be reconstructed correctly. And this is indeed guaranteed by {\it Protocol}~$2$, as all units CP$_{{\rm A}i}$ (CP$_{{\rm B}i'}$) obtain precisely the same information $p_{\rm A,info}$ ($p_{\rm B,info}$), $h_{\rm V}$ and $h_{\rm P}$. Finally, to reconstruct the key $k_{\rm A}$ and $k_{\rm B}$ from the shares $k_{{\rm A}ij}$ and $k_{{\rm B}i'j'}$, Alice and Bob now use the second step of the reconstruct protocol introduced in Appendix~\ref{ap_struct}.

Finally, one could modify {\it Protocol}~$3$ as follows. Like in the previous case, Alice and Bob replace the share, reconstruct, and RBS protocols with the protocols introduced above, given, of course, that the condition $P\notin\Sigma\sqcup\Omega\sqcup\Omega$ is fulfilled both in Alice and Bob. Also, in step~$1$, one obviously uses the version of {\it Protocol}~$2$ described in the previous paragraph. In addition, in step~$3$, one replaces the method to announce the hash function $h_{\rm P}$ with the procedure described in the previous paragraph. Moreover, also in step $3$, the parameter $t$ that appears in the expression of the secret key length now refers to the size of the biggest set in $\Sigma$. That is, $t$ is the maximum number of pairs of QKD modules that can be passively corrupted (or, alternatively, one can select $t$ as the maximum size of all maximal sets in a structure $\Sigma'$ that is obtained from $\Sigma$ by removing from all of its subsets those pairs of QKD modules which have aborted). Finally, Alice and Bob reconstruct $k_{\rm A}$ and $k_{B}$ by using again the second step of the reconstruct protocol introduced above.

Similar arguments can be applied as well to the protocol introduced in Appendix~\ref{alter} as alternative to {\it Protocol}~$2$.

\end{document}